\documentclass[twocolumn,floatfix,superscriptaddress,a4paper,showpacs,showkeys,nofootinbib,notitlepage]{revtex4-1}
\usepackage[colorlinks=true,linktocpage=true,linkcolor=blue,citecolor=blue,allcolors=blue]{hyperref}
\usepackage{epsfig}
\usepackage{latexsym}
\usepackage[utf8]{inputenc}
\usepackage{xspace}
\usepackage{indentfirst}
\usepackage{enumitem}
\usepackage{color}
\usepackage{placeins}

\usepackage{setspace}
\usepackage{lipsum}

\usepackage{hyperref}

\usepackage{todonotes}

\usepackage{amsmath}
\usepackage{amssymb}
\usepackage[english]{babel}
\usepackage{url}
\graphicspath{{figs/}}
\newcommand{\mean}[1]{\langle #1 \rangle}
\newcommand{\eq}[1]{\begin{align} #1 \end{align}}

\newcommand{\be}{\begin{equation}}
\newcommand{\ee}{\end{equation}}

\begin{document}

\title{Phase diagram of interacting pion matter and isospin charge fluctuations}

\author{O. S. Stashko}
\affiliation{
Taras Shevchenko National University of Kyiv, 03022 Kyiv, Ukraine
}
\author{O. V. Savchuk}
\affiliation{Frankfurt Institute for Advanced Studies, Giersch Science Center, D-60438 Frankfurt am Main, Germany}

\author{R. V. Poberezhnyuk}
\affiliation{Bogolyubov Institute for Theoretical Physics, 03680 Kyiv, Ukraine}

\author{V. Vovchenko}
\affiliation{Nuclear Science Division, Lawrence Berkeley National Laboratory, Berkeley, CA 94720, USA}

\author{M. I. Gorenstein}
\affiliation{Bogolyubov Institute for Theoretical Physics, 03680 Kyiv, Ukraine}

\date{\today}

\begin{abstract}
Equation of state and electric (isospin) charge fluctuations are studied for matter composed
of interacting pions.
The pion matter is described by self interacting scalar fields via a $\phi^4-\phi^6$ type Lagrangian. 
The mean-field approximation is used, and interaction parameters are fixed by fitting lattice QCD results on the isospin density as a function of the isospin chemical potential at zero temperature.
Two scenarios for fixing the model parameters -- with and without the first order phase transition -- are considered, both yielding a satisfactory description of the lattice data.
Thermodynamic functions and isospin charge fluctuations are studied and systematically compared for these two scenarios, yielding qualitative differences in the behavior of isospin charge susceptibilities.
These differences can be probed by lattice simulations at temperatures $T \lesssim 100$~MeV.

\end{abstract}
\maketitle

\section{Introduction}

The Bose-Einstein condensation (BEC)~\cite{Bose:1924mk,einstein1925stizunger} is 
a fascinating phenomenon that occurs in a system of bosons when a macroscopic amount of particles occupies the zero-momentum state.
This century-old phenomenon, observed experimentally in cold atomic gases~\cite{Anderson198,PhysRevLett.75.1687,PhysRevLett.75.3969,RevModPhys.71.463},
is predicted to occur in very different physical systems, ranging from condensed matter physics to high-energy nuclear physics, astrophysics, and cosmology~(see, e.g., Refs.~\cite{Satarov:2017jtu,Begun:2006gj,Begun:2008hq,Strinati_2018,Nozieres:1985zz,PhysRevLett.101.082502,Chavanis:2011cz,Mishustin:2019otg,Padilla_2019}). 
A theoretical description of the BEC appears to be rather sensitive to delicate details of particle interactions
\cite{kapusta_gale_2006,Andersen_2004,griffin1996bose,PhysRevA.88.053633,Watabe_2019,PhysRevLett.83.1703,Baym_2000, Holzmann_1999,Holzmann_2001,PhysRevLett.83.3770}.

In the present work we study the BEC phenomenon in strongly interacting QCD matter.
The effective low-energy degrees of freedom in QCD are pions -- the three pseudo-Goldstone bosons 
in the confined phase.
The pions obey the Bose-Einstein statistics, thus an emergence of the BEC of pions is possible and has been predicted to occur at large isospin chemical potentials, both in effective QCD theories~\cite{Son:2000xc,Abuki:2009hx} and in first-principle lattice QCD simulations~\cite{Brandt:2017oyy,Brandt:2018bwq}.
In nature, the pion BEC may occur during the cooling of the early Universe~\cite{Vovchenko:2020crk}, in the gravitationally bound pion stars~\cite{Brandt:2018bwq,Mannarelli_2019,andersen2018boseeinstein}, or as a non-equilibrium phenomenon in heavy-ion collisions~\cite{Begun:2006gj,Begun:2008hq,Begun:2015ifa}.
The hypothetical boson stars~\cite{Schunck_2003,2017Liebling,Braaten_2016} may exist
and can be a 
candidate for the dark matter in the Universe \cite{Su_rez_2013,Bernal_2017,Visinelli_2016,HajiSadeghi_2019,2011PhRvD..83d3525B,Gavrilik_2020}.

Different effective QCD descriptions of the phase diagram of interacting pion matter with a BEC include chiral perturbation theory~\cite{Adhikari:2019zaj,Adhikari:2020kdn}, Nambu-Jona-Lasinio model~\cite{He:2005nk}, Polyakov-loop extended quark meson model~\cite{Adhikari:2018cea,Folkestad:2018psc} etc.
Recently, a possibility of the BEC in the pion system at zero chemical potential was
considered within a Skyrme-like model including both attractive and repulsive interaction terms  
\cite{Mishustin:2019otg,Anchishkin_2019,stashko2020thermodynamic,stashko2020thermodynamic}.
Effects of  repulsive interactions on the BEC of pions were studied in Ref.~\cite{Savchuk:2020yxc} at non-zero chemical potential. 
The system of pions at zero chemical potential
was described~\cite{Mishustin:2019otg} by an effective Lagrangian with the attractive~($\phi^4$) and repulsive~($\phi^6$) terms of a scalar field $\phi$. 
In the present paper we extend this model 
to the  finite  
isospin  \footnote{We use the common simplified terminology and call the third component of isospin (electric charge) the isospin charge.}  chemical potential $\mu_I$.

The phase diagram on the whole plane of isospin chemical potential $\mu_I$ and temperature $T$ is investigated. 
Most macroscopic systems with both repulsive and attractive interactions between constituents display 
the first order liquid-gas phase transition (FOPT) which is ended by the critical point (CP). 
Therefore, these phenomena can  also be expected for the interacting pions 
in addition to the BEC. 

Lattice QCD results support an  existence of the pion BEC at finite isospin  chemical potential
~\cite{Brandt:2017oyy}.
We use the recent lattice data at zero temperature to fix the repulsive and attractive interaction parameters of the model. 
Then, thermodynamic functions and electric (isospin)
charge fluctuations up to the fourth order are calculated in the $(\mu_I,T)$-plane.
Two different scenarios are employed and systematically compared. 
The first one includes only the repulsive interactions via the $
\phi^6$ term, but not the attractive $\phi^4$ term.
In this case no FOPT transition is observed, only the BEC transition.
The second possibility takes into account both the repulsive and attractive pion-pion interactions. 
In this case the FOPT is observed at small $T$ and generates a non-trivial interplay between the FOPT and BEC transitions on the phase diagram.
The measures of the isospin charge fluctuations -- scaled variance, skewness, and kurtosis -- appear to be very sensitive to a presence of the CP and BEC phenomena. They are used to differentiate these two scenarios.

The paper is organized as follows: 
the theoretical description of interacting pion system is presented in Sec.~\ref{sec-formalism}. 
The two choices of the interaction potential, the fixing of the model parameters, and the resulting phase diagrams are discussed.
Section~\ref{sec-fluks} is dedicated to fluctuations of the isospin charge, in particular the  scaled variance, skewness, and kurtosis are discussed in some detail.
Summary in Sec.~\ref{sec-conc} closes the paper.

\section{Interacting pion system}
\label{sec-formalism}

\subsection{Model formulation}

The three pions species, $(\pi^+,\pi^-,\pi^0)$, are represented as a triplet of interacting  pseudo-scalar fields $\phi=(\phi_1,\phi_2,\phi_3)$ that are described by an effective relativistic Lagrangian density:
\eq{
{\cal L}=\frac{1}{2}\left(\partial_\mu\phi\,\partial^{\,\mu}\phi -m_{\pi}^2 \phi^2\right)+{\cal L}_{\rm int}\left(\phi^2\right)~,
}
where $m_{\pi}$ is the vacuum pion mass\footnote{We use the natural units, $\hbar = c = k = 1$, and assume equal masses of all three pion species, $m_{\pi}=140$~MeV.} and  ${\mathcal L}_{\rm int}$ is the interaction part of the Lagrangian. 
We omit here the electromagnetic interactions.
Consider now this system in statistical equilibrium within the grand canonical ensemble (GCE).
The independent variables are the temperature $T$ and the isospin chemical potential $\mu_I$.
The isospin chemical potential couples to the conserved isospin  charge, the pion species $\pi^+$, $\pi^-$, and $\pi^0$ carry the isospin charges of +1, -1, and 0, respectively.

To proceed, we apply a relativistic mean-field approximation, i.e. series ${\mathcal L}_{\rm int}$ in terms of $\delta\sigma=\phi^2-\sigma$, where 
$\sigma=\langle\phi^2\rangle$ is the expectation value of the scalar field and $\mean{...}$ denotes
the GCE  averaging.
The effective mean-field Lagrangian 
can then be represented as \cite{Mishustin:2019otg}:
\eq{\label{lag}
{\cal L}\approx\frac{1}{2}\left[\partial_\mu\,\phi\,\partial^{\,\mu}\phi -M^2(\sigma) \phi^2\right]
+p_{\rm ex}(\sigma)\,,
}
where $M(\sigma)$ is the effective pion mass 
and $p_{\rm ex}(\sigma)$ is the so-called excess pressure,
\eq{\label{eq:Mpex}
M^2(\sigma)=m_{\pi}^2-2\frac{d{\cal L}_{\rm int}}{d\sigma}~,~~~p_{\rm ex}(\sigma)=
{\cal L}_{\rm int}-\sigma \frac{d{\cal L}_{\rm int}}{d\sigma}.}
The effective Lagrangian form of Eq.~\eqref{lag} implies that the main effect of interactions in our description leads to an appearance of a medium-dependent effective mass $M(\sigma)$.
The excess pressure $p_{\rm ex}(\sigma)$ -- the second term in the right hand side  of Eq.~\eqref{lag} -- ensures the proper counting of the interaction energy.

The details of the model formulation can be found in Refs.~\cite{Mishustin:2019otg,Satarov:2020loq}. This model  was previously used to describe the pion system at zero chemical potential \cite{Mishustin:2019otg} and the system of interacting alpha particles \cite{Satarov:2020loq}. 
In the present study we  apply this model to the new physical situation and consider the pion system 
at non-negative values of the isospin chemical potential
$\mu_I \ge 0$. 
The results at $\mu_I \le 0$ can then be obtained by interchanging $\pi^+$ and $\pi^-$.
Values of $\mu_I > 0$ correspond to positive values of the isospin charge density $n_I \equiv n_+-n_- >  0$, where $n_+$ and $n_-$ correspond to $\pi^+$ and $\pi^-$ particle number densities, respectively. 
We consider the possible BEC of the positively charged pions in this regime.
The expectation value of the scalar field $\sigma$ is presented as
(see Refs.~\cite{Mishustin:2019otg,Satarov:2020loq} for the derivation details)
\eq{\label{sigma}
\sigma(T,\mu_I,M)= \sum_{i}\sigma_i^{\rm th}(T,\mu_i,M)~+~\sigma_+^{\rm bc}~,
}
where $\sigma^{\rm th}_i$ correspond to the contributions of the thermal pions,
$ i=(+,-,0~)$,
\eq{\label{sigma-th}
\sigma_i^{\rm th}(T,\mu_i,M) =\int \frac{d^3 k}{(2 \pi)^3} \frac{n_{\rm k}(T,\mu_i,M)}{\sqrt{k^2+M^2}}~, 
}
while $\sigma_+^{\rm bc}$ corresponds to a possible contribution of the Bose condensate (BC) of $\pi^+$. 
Here $\mu_+=\mu_I$, $\mu_-=-\mu_I$, $\mu_0=0$, 
and 
\eq{\label{nk}
n_{\rm k}(T,\mu_i,M)=\left[\exp\left(\frac{\sqrt{k^2+M^2}-\mu_i}{T}\right)-1\right]^{-1}\,.
}

We will use a Skyrme-like parameterization of the interaction term:
\eq{\label{Lint}
\mathcal{L}_{\rm int}(\sigma)=\frac{a}{4} \sigma^2-\frac{b}{6} \sigma^3\,,~~~~a\ge 0,~b > 0~.
}
In Eq.~(\ref{Lint}), $a\ge 0$ and $b>  0$ are model parameters which define the strength of, respectively, attractive and repulsive interactions between particles.
The effective mass and the excess pressure for this choice of the interaction terms 
are equal to:
\eq{
M(\sigma)&=\sqrt{m_{\pi}^2-a \sigma +b \sigma^2}\,,\label{mass}\\
p_{\,\rm ex}(\sigma)&=
-\frac{a}{4} \sigma^2 +\frac{b}{3} \sigma^3\,.\label{epre2}
}
Inverting Eq.~(\ref{mass}) with respect to $\sigma$ we obtain\footnote{
The second root of Eq.~\eqref{mass} corresponds to mechanically unstable states.}
\eq{\label{s-M}
\sigma= \frac{a+\sqrt{a^2+4~b~( M^2-m_{\pi}^2)}}{2~b}.
}
At given $T$ and $\mu_I$ we use the system of self-consistent Eqs.~(\ref{sigma}) and (\ref{mass}) to determine $\sigma$ and $M$.

The  pressure~$p$,  the number densities of thermal pions $n_i^{\rm th}$, and the isospin charge density~$n_I$ can be calculated as 
\eq{
\label{p}
& p = \sum _i \int \frac{d^3 k}{(2 \pi)^3}\frac{k^2}{\sqrt{k^2+M^2}} n_{\rm k}(T,\mu_i,M)
+ p_{\rm ex}(\sigma),\\
\label{eq:ni}
& n_i^{\rm th}  = \int \frac{d^3 k}{(2 \pi)^3}n_{\rm k}(T,\mu_i,M), \\
\label{ni}
& n_I = \left(\frac{\partial p}{\partial \mu_I}\right)_T~=~n_+-n_-~. 
}
Here $n_i$ are the total number densities of
pions. The $n_+$ density may include a contribution from a Bose condensate (BC).
The condensation does not occur if $\mu_I<M$.
In the case $\mu_I<M$ only the thermal pions contribute to the total number densities,
\eq{
n_+=n_+^{\rm th}~,~~~~n_-=n_-^{\rm th}~,~~~~n_0=n_0^{\rm th}~, \quad \mu_I < M~.
}

The BEC of $\pi^+$ occurs when their chemical potential $\mu_+ \equiv \mu_I$ reaches the value of the effective mass,\footnote{Note that chemical potential values $\mu_I > M$ exceeding the effective mass are forbidden as they would lead to negative occupancy numbers $n_{\rm k}$ (\ref{nk}) for some $k$-states. } i.e. $\mu_I = M$.
In this case the number density $n_+$ may receives a contribution $n_+^{\rm bc}$ from the 
BC:
\eq{
n_+ = n_+^{\rm th}+ n_+^{\rm bc}~, \qquad \mu_I = M.
}
The number of densities of the two other pions species are unchanged: $n_- = n_-^{\rm th}$, $n_0 = n_0^{\rm th}$.
The number density of $\pi^+$ in a condensate reads
\eq{\label{sigma+}
n_+^{\rm bc}~=~\mu_I~\sigma_+^{\rm bc}~=~\mu_I~\left(\sigma~-~\sum_i\sigma_i^{\rm th}\right)~.
}
In Eq.~(\ref{sigma+}) the quantities $\sigma$ and $\sigma_i^{\rm th}$ are calculated according to Eq.~(\ref{s-M}) and Eq.~(\ref{sigma-th}), respectively.

An  onset of the BEC takes place when $\mu_I$ reaches the effective mass $M$.
This condition defines a line in the phase diagram -- the BEC line.
This line is calculated by substituting $M \to \mu_I$ and $\sigma_+^{\rm bc} \to 0$ into the system of equations~(\ref{sigma}),~(\ref{mass}), and (\ref{eq:ni}) and solving it with respect to $\mu_I$ at given value of $T$.

\subsection{Fixing the parameters using lattice data at zero temperature}

Lattice QCD simulations at finite isospin density provide constraints on the equation of state from first principles~\cite{Brandt:2017oyy}.
In particular, the isospin density $n_I(T=0,\mu_I)$ at zero temperature has been presented in~\cite{Brandt:2018bwq}.
Here we will use these lattice data to constrain the parameters of our model.

\begin{figure*}
         \includegraphics[width=0.49\textwidth]{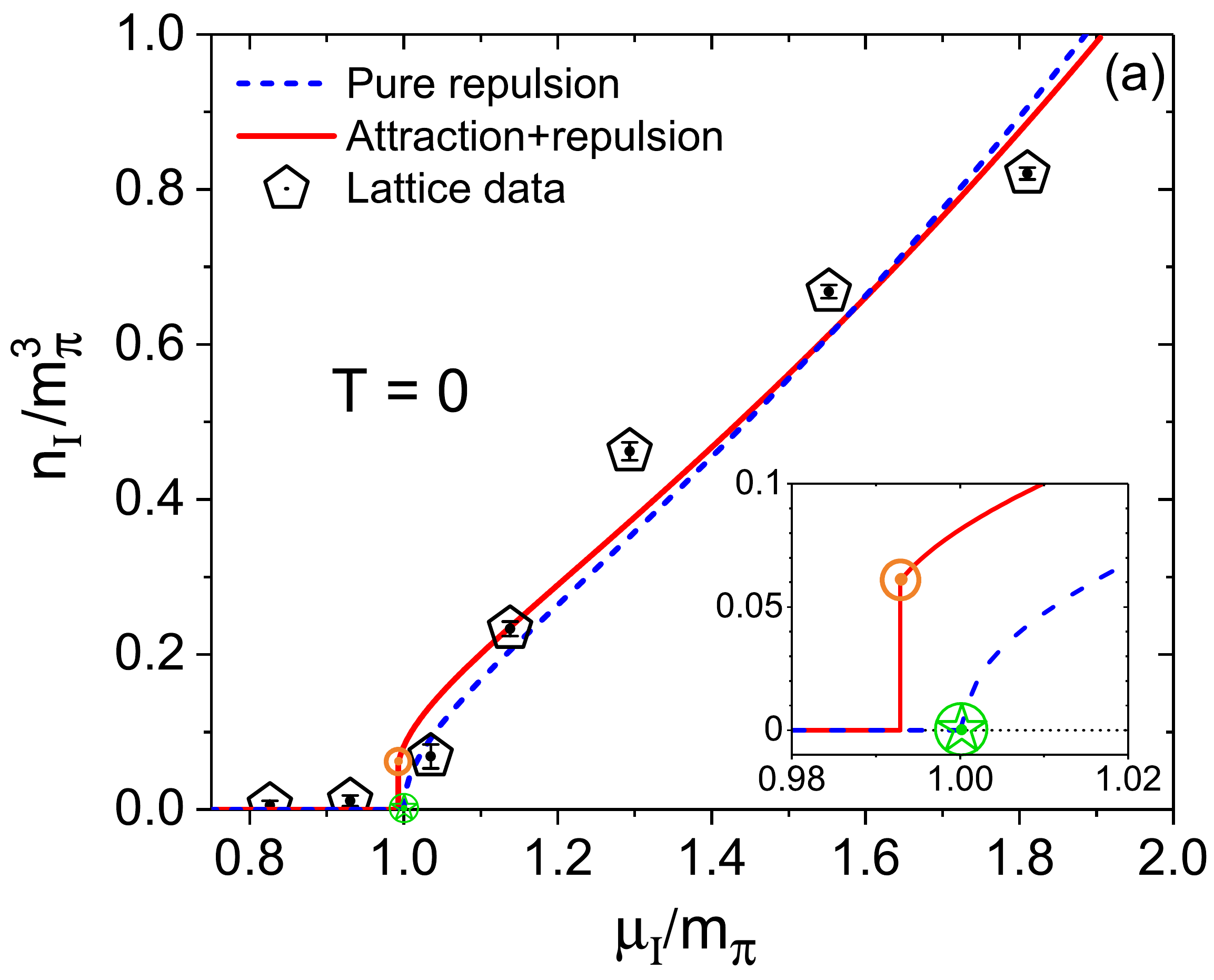}
         \includegraphics[width=0.49\textwidth]{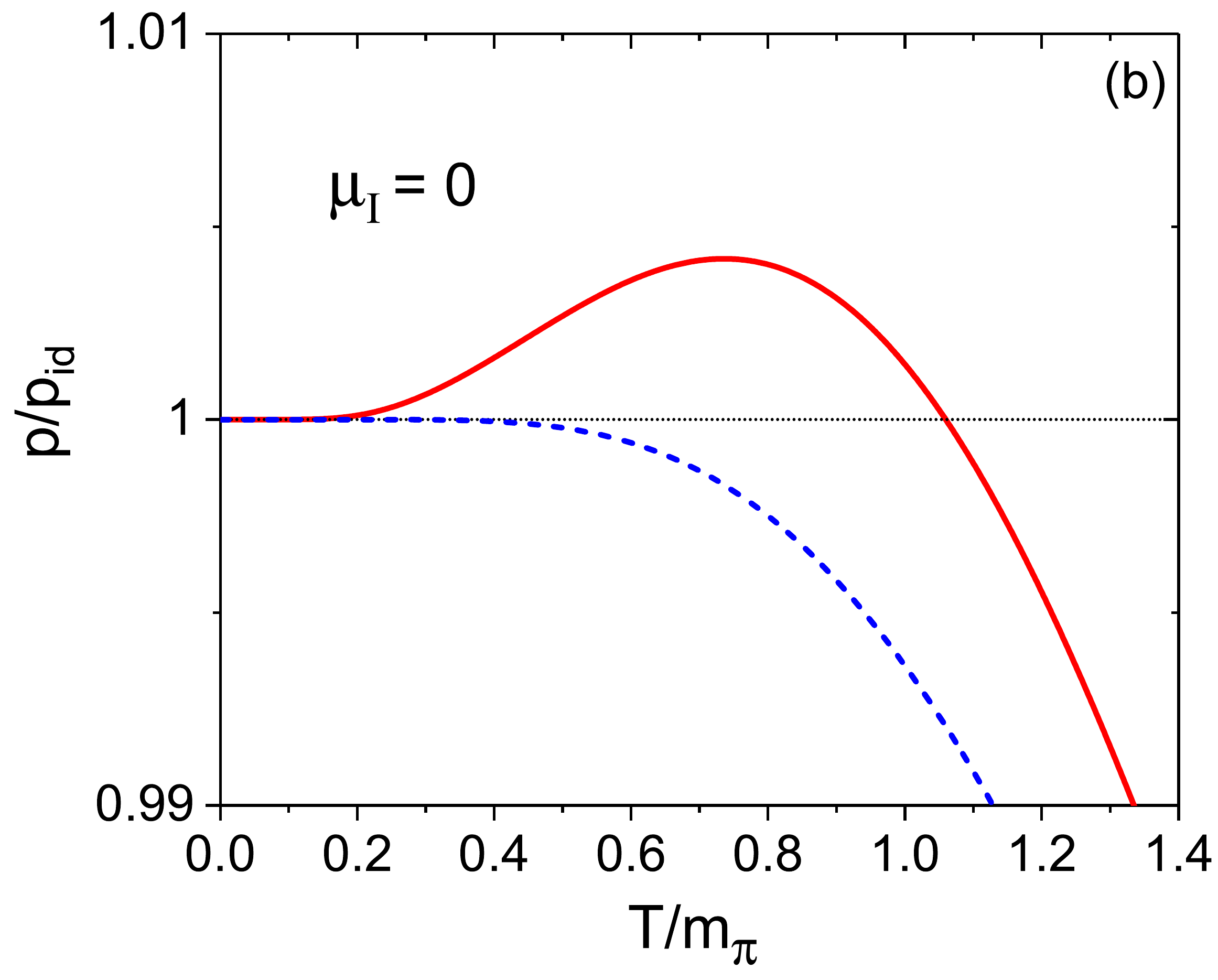}
     \caption{(a): A comparison of the  pion condensate as a function of the isospin chemical potential, $\mu_I$, at zero temperature, $T=0$, with lattice data of Ref.~\cite{Brandt:2018bwq} is shown by solid line and dashed line for the Scenario I ($a=0$) and Scenario II ($a>0$), respectively.
     The ground state $n_I\approx 0.022~{\rm fm}^{-3}$ at $\mu_I\approx 0.993~m_\pi$ for  $a>0$ and the singularity point $n_I=0$ at $\mu_I=m_\pi$ for $a=0$ are marked by, respectively, the red circle and the hollow green star. The inset shows the zoomed in picture close to $\mu_I=m_{\pi}$. 
      (b): The ratio of the system's pressure to the ideal gas pressure at zero isospin chemical potential, $\mu_I=0$, as a function of temperature. 
     }
    \label{fig-fit}
\end{figure*}

In the limit of zero temperature, $T = 0$, the thermal pion excitations are absent, i.e. all thermal densities (\ref{eq:ni}) vanish. 
In this case the system consists solely of the BC of $\pi^+$-mesons, thus, the isospin density coincides with the number density of the condensed pions, $n_I(T=0,\mu_I) = n^{\rm bc}_+$, and the total pressure equals to the excess pressure, $p(T=0,\mu_I) = p_{\rm ex}$.
The explicit expression for $n_I(T=0,\mu_I)$ in the the considered model follows from Eqs.~(\ref{sigma-th}), (\ref{s-M}), and \eqref{sigma+}:
\eq{\label{n0}
n_I(T=0,\mu_I) = \mu_I\left(\frac{a+\sqrt{a^2+4~b~( \mu_I^2-m_{\pi}^2)}}{2~b}\right).
}

\subsubsection{Scenario I: Repulsive interactions only}

In the first scenario we consider purely repulsive interactions between pions.
To achieve this we set $a = 0$ and $b > 0$.\footnote{Another option would be to set $b = 0$ and take $a < 0$. The results in such a case are qualitatively similar to $a = 0$ and $b > 0$.}
Equation~(\ref{n0}) in this case is reduced to
\eq{\label{a0}
n_I(T=0,\mu_I) =  b^{-1/2}\,\mu_I \sqrt{\mu_I^2-m_{\pi}^2}~\theta(\mu_I-m_{\pi})~.
}
An onset of the BEC occurs at $\mu_I = m_{\pi}$.
The isospin density is a continuous function of $\mu_I$ since $n_I(T=0,\mu_I = m_{\pi}) = 0$. 
On the other hand, the $\mu_I$-derivative of $n_I$ exhibits a discontinuity at $\mu_I = m_{\pi}$.
Therefore, the transition between vaccuum and a pion-condensed phase at $\mu_I = m_{\pi}$ is a second-order phase transition at $T = 0$.
Qualitatively, this is consistent with predictions of many different theories, including for instance chiral perturbation theory~\cite{Son:2000xc,Adhikari:2019zaj} or Polyakov-loop extended quark meson model~\cite{Adhikari:2018cea,Folkestad:2018psc}.

To fix the value of the parameter $b$ we fit the lattice QCD data on $n_I(T=0,\mu_I)$ of Ref.~\cite{Brandt:2018bwq} in the range of chemical potentials~$\mu_I/m_{\pi} < 2$.
We obtain $b \simeq 9.09/m_{\pi}^2$ with $\chi^2 / dof \simeq 1.62$.
A comparison with the lattice data is shown in Fig.~\ref{fig-fit} (a) by  blue dashed line.

\subsubsection{Scenario II: Repulsion + attraction}

Let us turn now to the more general case when both the attractive and repulsive interactions are present, $a > 0$ and $b > 0$.
In this case the system undergoes a first-order phase transition between vacuum~(the {\it gaseous} phase) and a pion-condensed phase~(the {\it liquid} phase), with the coexistence point being characterized by a vanishing pressure.
To see this consider Eq.~\eqref{epre2}: 
this equation have two solutions for $p_{\rm ex} = p(T=0,\mu_I) = 0$,
defining the expectation values of the scalar field at the FOPT boundaries,
$\sigma_g = 0$ and $\sigma_l =3a/4b$.
This corresponds, via Eq.~(\ref{mass}), to the following values
of the effective mass in the gaseous and liquid components:
\eq{\label{sM}
M_g~=~m_{\pi}~,~~~~M_l~=~ \sqrt{m_{\pi}^2-\frac{3a^2}{16b}}~.
}

The FOPT takes place at $\mu_0 = M_l$. 
The gaseous phase at $T = 0$ corresponds to the vacuum, thus, $n_g = 0$.
The isospin density jumps at $\mu_I = \mu_0$ from $n_g = 0$ to
\eq{
n_l \equiv n_I(T=0,\mu_I \to \mu_0 + 0)=\frac{3a}{4b}\, \sqrt{m_{\pi}^2-\frac{3a^2}{16b}}.
}

To fix the numerical values of $a$ and $b$ we again fit the lattice data on $n_I$ at $T = 0$.
We obtain $a \simeq 0.93$ and $b \simeq 11.39 / m_{\pi}^2$ with a fit quality of $\chi^2/dof \simeq 1.35$ -- a slightly better fit compared to Scenario I.

As discussed above, Scenario II predicts the FOPT.
Using the numerical values of the $a$ and $b$ parameters fitted to the lattice data,
one obtains the FOPT at $\mu_0 \approx 0.993\,m_{\pi}$, where the isospin density jumps from $n_g = 0$ to $n_l >0$.
The values of the $\pi^+$ density $n_I$ and the binding  energy per particle
$W$ at $T=0$ and $\mu_I=\mu_0$, 
\eq{\label{GS}
& n_I \approx 0.060~m_\pi^3\approx  0.022~{\rm fm}^{-3}~,\\
& W \equiv \frac{\varepsilon}{n_I} - m_\pi= M_l-m_{\pi}\approx -~1~{\rm MeV}~,\label{W}
}
obtained in Scenario II correspond to the ground state of the pion matter.
This density is about 7 times smaller than the normal nuclear matter density of $n_0 = 0.16$~fm$^{-3}$, and the binding energy is about 16 times smaller than that in the nuclear ground state.

The behavior of $n_I$ in Scenario II at zero temperature is shown in Fig.~\ref{fig-fit} (a) by the solid red line.
A comparison with the lattice data and the predictions of Scenario I are also shown.
Overall, the behavior of $n_I$ in the both scenarios is similar.
Even though a nature of the phase transition differs between the two scenarios, due to the small latent heat of the FOPT in Scenario II it is difficult to distinguish it from the second-order phase transition in Scenario I using the presently available lattice data.
In Sec.~\ref{sec-fluks} we discuss fluctuations as a possibility to make such a distinction.

\begin{figure*}
\includegraphics[width=\textwidth]{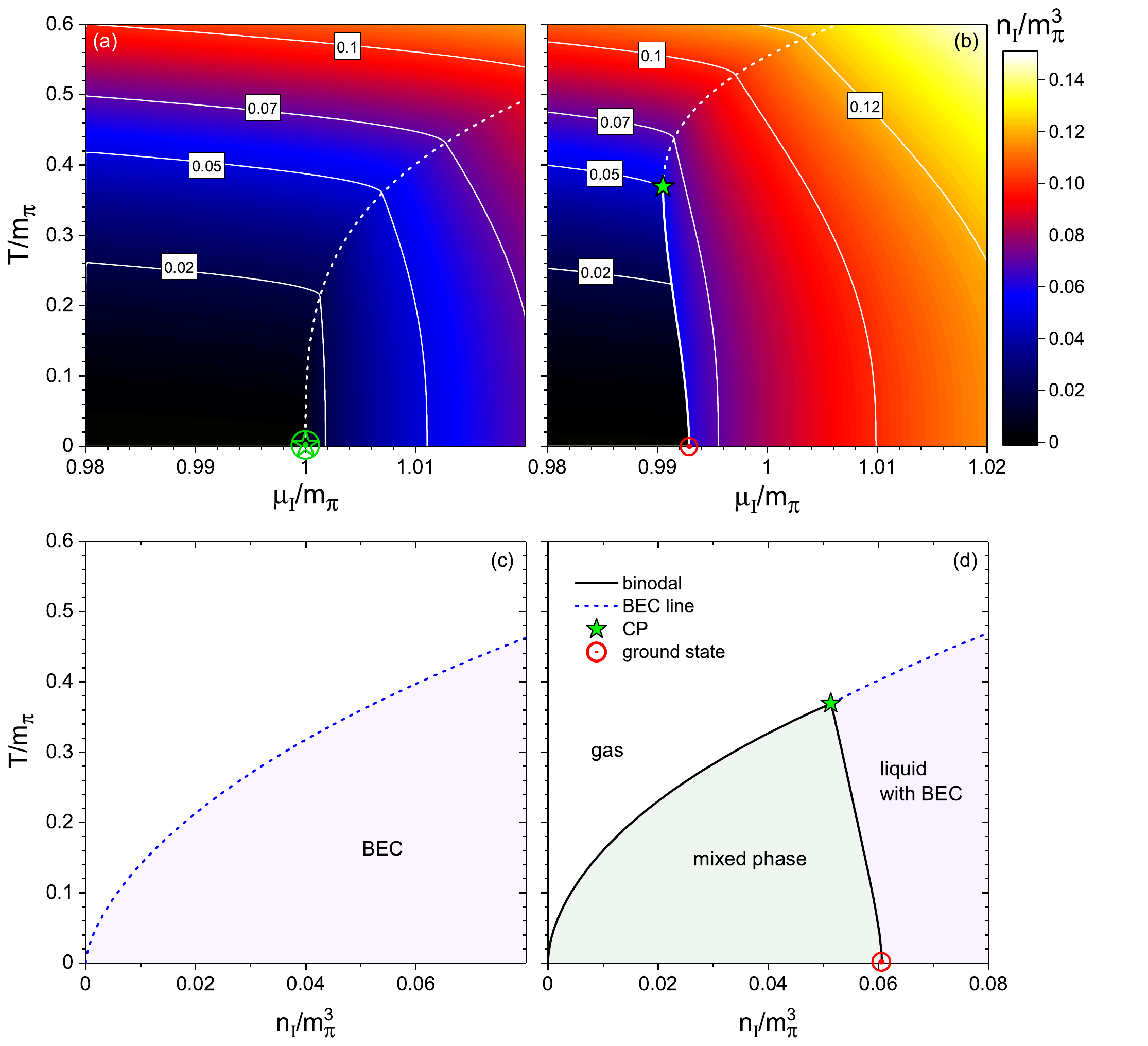} 
     \caption{Phase diagrams of the pion matter in the $(\mu_I/m_{\pi}, T/m_{\pi})$  and ($n_I/m_{\pi}^3,T/m_{\pi})$ planes, 
     at $a=0$ [(a) and (c)], and  $a>0$ [(b) and (d)].
     Dashed lines correspond to the onset of the BEC.
     Solid lines correspond to the first order phase transition.
     The CP  is marked by the green star.
     Coloring in pallets (b) show the isospin charge density, $n_I/m_{\pi}^3$.
     }
    \label{fig-diagram}
\end{figure*}

\subsection{Phase diagram at finite temperatures}

Model calculations at finite temperatures are straightforward.
Important constraints on the equation of state of pion matter can be obtained at zero chemical potentials and large temperatures, $120 \leq T \leq 160$~MeV, where the QCD equation of state is known from lattice QCD~\cite{Borsanyi:2013bia,Bazavov:2014pvz}.
In this range, the pressure and energy density are reasonably well described by the ideal hadron-resonance gas (see e.g., Ref.~\cite{Vovchenko:2014pka}). 
This indicates that effects of pion interactions in this regime are small.
To verify this we plot in Fig.~(b) a ratio of the pressure of interacting pions to the ideal pion gas baseline~(i.e., at $a=0$ and $b=0$) for the two scenarios.
For purely repulsive interactions~(scenario I) the pressure demonstrates small suppression relative to the ideal gas.
If attractive effects are included~(scenario II) the pressure at small  temperatures 
is higher than that of the ideal gas of pions. However, at large $T$ the repulsive effects become dominant and the pion pressure is again suppressed.
In both scenarios the corrections to the ideal gas pressure at $\mu_I = 0$ are small, not exceeding 1\%.
This is not the case at non-zero values of $\mu_I$: the thermodynamics of the interacting pion gas
differs drastically from that in the ideal pion gas in the $(\mu_I,T)$-region of the phase diagram where a BEC pions is formed, as discussed in the following.

In Scenario I, where the attractive pion interactions are absent ($a=0$), there is no FOPT 
in the pion system.
An onset of the $\pi^+$ BEC takes place when $\mu_I$ reaches the value of the effective mass $M$. 
The BEC line can thus be obtained by substituting $M \to \mu_I$ in the system of equations~(\ref{sigma}) and (\ref{mass}) and solving it with respect to $T$. 
The resulting BEC line $T_{\rm bc}(\mu_I)$ is shown in Fig.~\ref{fig-diagram} (a) by the dashed line. 
In the ideal gas limit one would obtain a vertical BEC line, $\mu_I=m_{\pi}$.
The deviation from the ideal gas behavior thus becomes evident as $T$ is increased.
This is due to large particle number densities, and thus stronger effects of interactions, as the temperature is increased.
Note that in the ideal Bose gas, a region of the $(\mu_I,T)$-plane with $\mu_I>m_{\pi}$ is forbidden, whereas in the interacting system considered here this region 
is legitimate.
It follows from Eq.~(\ref{s-M}) that the effective mass is always larger than the vacuum mass, $M(T,\mu_I)>m_{\pi}$, the pure repulsion scenario ($a=0$).
The $(n_I,T)$ phase diagram in the $a=0$ scenario is shown in Fig.~\ref{fig-diagram} (c).
The thermodynamic states below the dashed lines in Figs.~\ref{fig-diagram} (a) and (c) correspond to a non-zero density of the BC, i.e. to a macroscopic number of $\pi^+$-mesons occupying the zero momentum level $k=0$.

In Scenario II, with both the repulsive ($b>0$) and attractive ($a > 0$) pion interactions present, the FOPT  phase transition takes place in addition to the BEC formation.
The $(\mu_I,T)$ and $(n_I,T)$ planes are presented for this scenario in  Figs.~\ref{fig-diagram} (b) and (d), respectively.
The line of the FOPT is shown by a thick solid line in Fig.~\ref{fig-diagram} (b). 
This line ends in a critical point (CP) at
$T=T_c\approx 0.369~m_{\pi}$, $\mu_I=\mu_c\approx 0.991~m_{\pi}$, and $n_I=n_c\approx 0.051~m_{\pi}^3$, which is shown by the green star. 
The $\mu_c$ value can be expressed explicitly in terms of model parameters
\eq{ \label{muc}
\mu_{c} ~=~ \sqrt{m_\pi^2 - \frac{a^2}{4b}}.
}
An approximate analytical dependence of $T_c$ on $a,b$, and $m_{\pi}$ can also be obtained 
\eq{\label{Tc}
T_c \approx ~\frac{\pi}{\sqrt{b}}\left(\frac{2 a}{\zeta(3/2)}\right)^{2/3}\left(4bm_{\pi}^2-a^2\right)^{-1/6}.
}
Here $\zeta(x)$ is the Riemann zeta function. 
The relation (\ref{Tc}) has been obtained assuming 
$n_-\ll n_0\ll n_+$ as well as 
the non-relativistic approximation in vicinity of the CP.
Using the previously obtained parameters $a$ and $b$ from fitting the lattice data one obtains $T_c\approx 0.393\, m_{\pi}$. 
This is within 6\% of the numerical result obtained without approximations.
Note that the limit $a\rightarrow 0$ corresponds to $T_c\rightarrow 0$ and $\mu_c\rightarrow m_{\pi}$.

At the FOPT line in the $(\mu_I,T)$-plane the pressures of the {\it gaseous} and {\it liquid} phases are equal to each other. 
On the other hand, the isospin charge density $n_I$ has a discontinuity. 
The mixed phase shown in Fig.~\ref{fig-diagram} (d) is bounded by the gas-like (left) and liquid-like (right) binodals presented by solid lines that intersect each other at the CP.  
The pion states inside the mixed phase correspond to linear combinations of the diluted (gaseous) and dense (liquid) states lying on the left and right binodals, respectively. 
The liquid component of the mixed phase always lies below the BEC line, thus it always contains a non-zero fraction of condensed $\pi^+$-mesons.
The gaseous component, on the other hand, does not contain the BEC.

A remarkable feature of the considered model is that the BEC line enters the mixed phase at the CP. 
This property of the model is robust with regard to variations in the values of the $a$ and $b$ parameters. 
Another peculiar property is the non-smooth intersection of the left and right binodals at the CP.

Scenarios I and II provide a similar picture of the phase diagram at
$T\gg T_{c}$.
At $T \lesssim T_c$, on the other hand, the differences are significant. 
We argue that these differences can be most clearly seen by studying the behavior of isospin charge fluctuations.
This is discussed in Sec.~\ref{sec-fluks}.

\begin{figure*}
      \includegraphics[width=\textwidth]{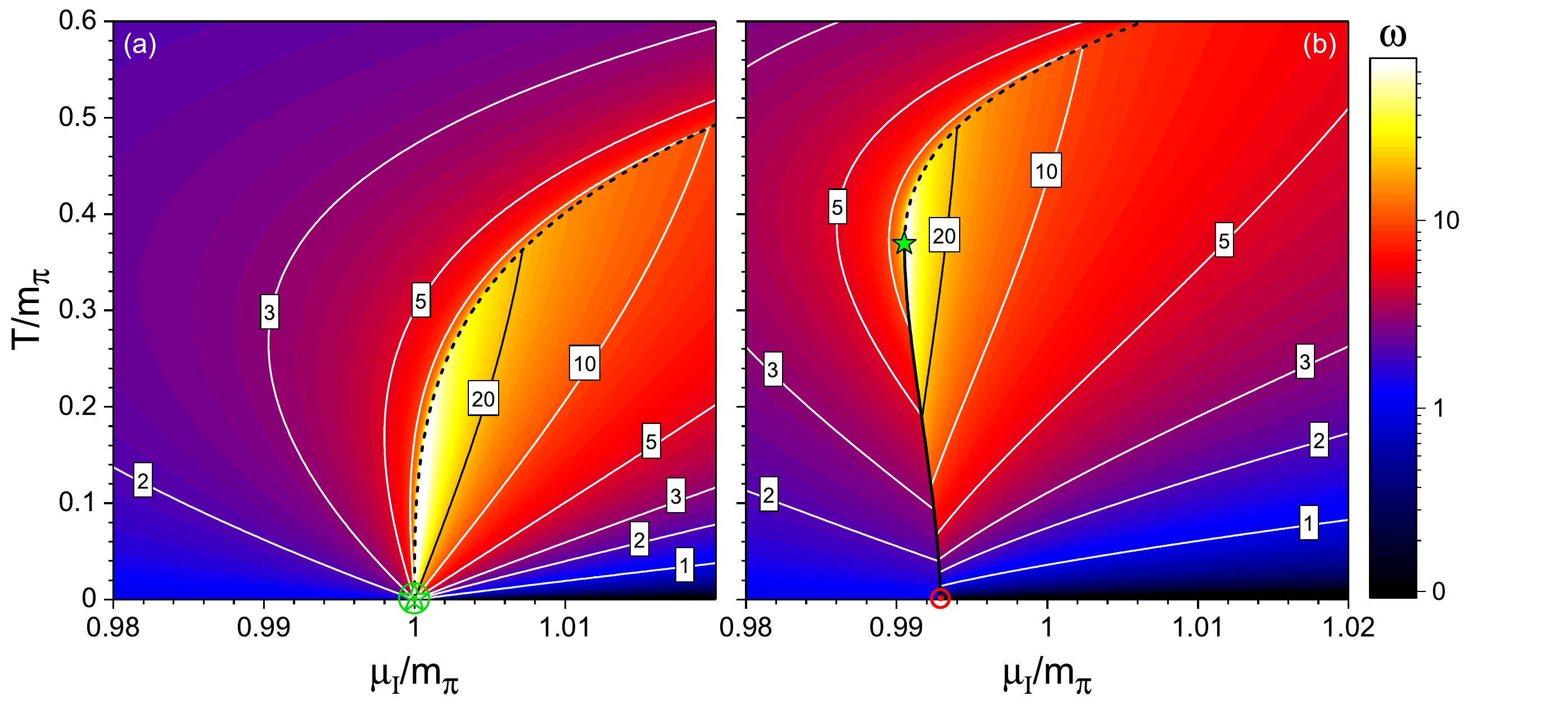} 
    \caption{The scaled variance of the isospin charge fluctuations of the pion matter in the $(\mu_I/m_{\pi}, T/m_{\pi})$ plane is shown for (a) the pure repulsion case (Scenario I) and (b) the full potential case (Scenario II). 
     Colors show values of the scaled variance. Dashed lines correspond to the onset of the BEC.
     Black solid lines correspond to the first order phase transition.
     The CP is marked by the green star. \label{fig-omega}  }
\end{figure*}

\section{Fluctuations}
\label{sec-fluks}

The two presented descriptions of the isospin charge density from the lattice results at $T=0$
both contain the BEC. Within the second description, the FOPT at $T<T_c$ 
leads to the $n_I$ discontinuity. 
We argue that the difference between the two scenarios can be probed by considering isospin charge fluctuations.

\begin{figure*}
         \includegraphics[width=\textwidth]{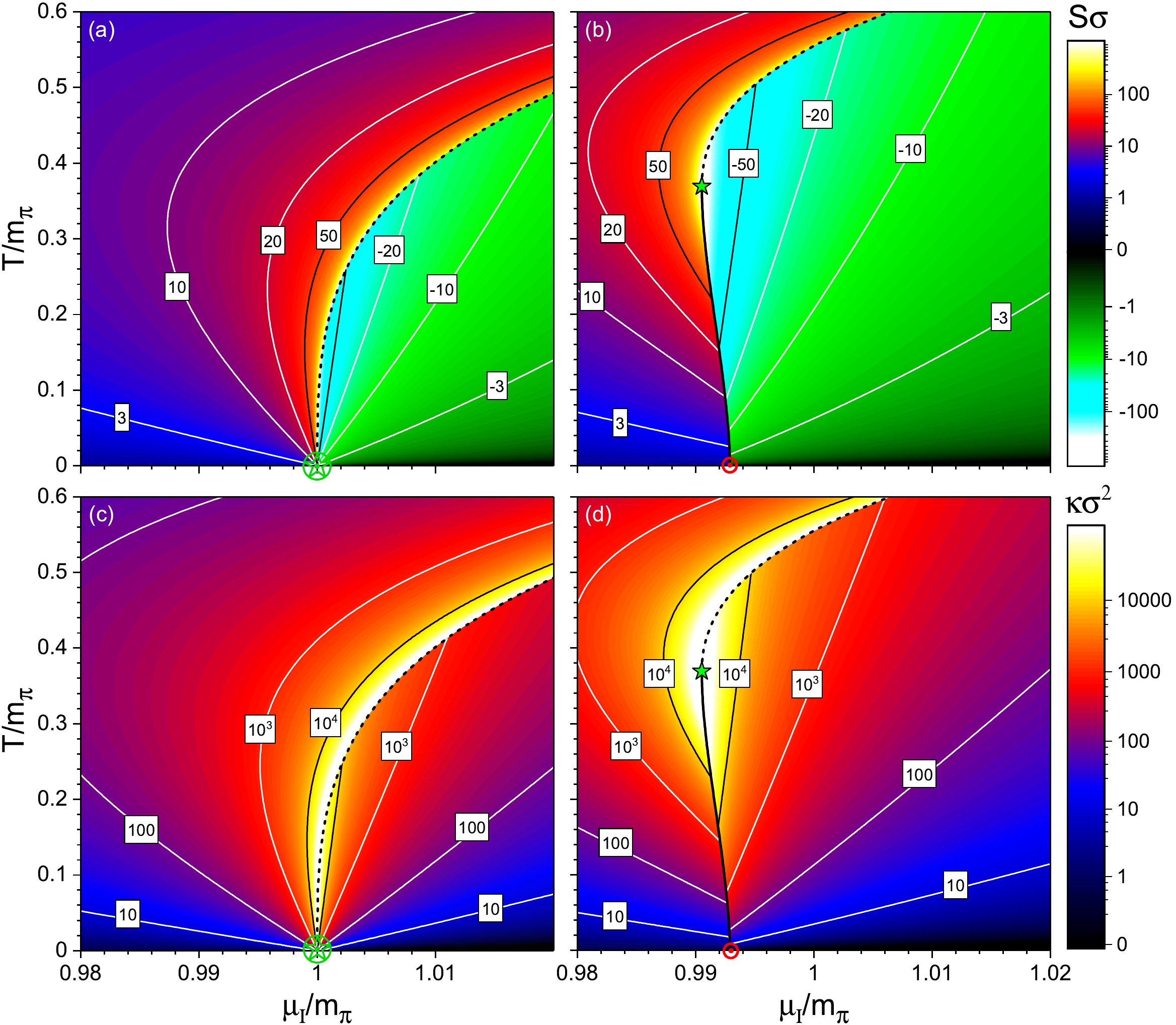}
        \caption{The skewness (a),(b) and kurtosis (c),(d) of the isospin charge fluctuations  in $(\mu_I/m_{\pi},T/m_{\pi})$ are shown for (a),(c) the pure repulsion case (Scenario I) and (b),(d) the full potential case (Scenario II). \label{fig-skewness}}
\end{figure*}

In the GCE, the $j$-th order
susceptibility of the isospin charge 
is determined by a $j$-th order partial derivative of the pressure $p$ with respect to the chemical potential $\mu_I$:
\eq{\label{cum}
\chi_j ~=
~\frac{\partial^j (p/T^4)}{\partial (\mu_I/T)^ j}~.
}
Ratios of susceptibilities given by (\ref{cum}) can be particularly useful as such quantities are intensive in the thermodynamic limit.
Some of the most well known such quantities include
the scaled variance $\omega$, 
skewness $S\sigma$, and kurtosis $\kappa\sigma^{2}$ (see, e.g., Ref.~\cite{Karsch:2010ck}):
\eq{\label{cumB}
\omega =\frac{\chi_2}{\chi_1},~~~~~  S\sigma=\frac{\chi_3}{\chi_2},~~~~~ \kappa\sigma^{2}=\frac{\chi_4}{\chi_2}.
}

Using Eq.~(\ref{cum}) together with Eq.~(\ref{p}) the scaled variance  $\omega$ can be written as
\eq{
\omega=\frac{T}{n_I}\left(\frac{\partial n_I}{\partial \mu_I}\right)_T.
}
In the ideal pion gas the scaled variance diverges at the BEC line \cite{Begun:2006gj}, i.e., $\omega_{\rm id}\rightarrow \infty$ at $\mu_I\rightarrow m_{\pi}-0$:
\eq{\label{id}
\omega_{\rm id}= T^2\frac{m_{\pi}^{3/2}}{\sqrt{2}\pi n_I} (m_{\pi}-\mu_I)^{-1/2}\rightarrow{\infty}~.
}
Due to the repulsive interactions in the considered model the scaled variance remains finite. On the BEC line, 
$M=\mu_I$, one finds:  
\eq{\label{omega-bec-line}
\omega = \frac{MT}{n_I}\left(\frac{\partial M}{\partial \sigma}\right)^{-1} 
= \frac{\mu_I^2T}{n_I}\frac{1}{\sqrt{b(\mu_I^2-\mu_{c}^2)}},~
}
where $\mu_{c}$ is 
given by Eq.~(\ref{muc}).
The value of $\omega$ remains also finite in a presence of the BC, $n_+^{\rm bc}>0$. 
In Scenario II~($a > 0$), $\omega$ exhibits singular behavior at the CP,  $T=T_c$, $\mu_I=\mu_c$, where it diverges.
A systematic expansion of the thermodynamic functions in a vicinity of the CP
allows to obtain the critical exponents. 
We expect that the critical exponents of the considered system are different from those in the  mean-field class universality. 
This is  due to a presence of the two order parameters, $n_I^{l}-n_I^{g}>0$ and $n_+^{\rm bc}>0$,
which disappear simultaneously at the CP (see, e.g., Ref.~\cite{PhysRevE.79.021116}). 
A detailed discussion of this subject is however
outside of the scope of the present study.

\paragraph*{Scaled variance.} 
The behavior of the scaled variance $\omega$ in the plane of temperature and isospin chemical potential is shown in Fig.~\ref{fig-omega}. 
In Scenario~I~(pure repulsion), $\omega$ is a continuous function, in particular across the BEC boundary~[see~Fig.~\ref{fig-omega}~(a)].
In Scenario II~(full potential), on the other hand,
$\omega$ exhibits a jump discontinuity over the FOPT line an becomes divergent at the CP~[Fig.~\ref{fig-omega}~(b)].
It is still a continuous function across the BEC line, however.

Note that in the limit $a \rightarrow 0$~(Scenario I), 
the CP approaches $T_c\rightarrow 0$ and $\mu_c\rightarrow m_{\pi}$.
Therefore, the point $\mu_I=m_{\pi}$ at zero temperature retains some of the proper tires of the CP and exhibits large fluctuations in its vicinity.
One can observe $\omega$ of any magnitude in the vicinity of this point, the exact magnitude depending on the path of approach.
In particular, approaching this point along the BEC line one finds $\omega\rightarrow \infty$ at $T\rightarrow 0$.

\paragraph*{Skewness.}
The skewness, $S\sigma$, for Scenario I~($a=0$) and II~($a>0$) is shown in Figs.~\ref{fig-skewness} (a) and (b), respectively.
At small $\mu_I\ll m_{\pi}$ values, where the pion densities are small,
both the pion interactions and Bose statistics effects can be neglected, thus, $S\sigma \approx 1$.
The skewness attains positive values in those regions of the phase diagram where there is no BC.
$S\sigma$ is discontinuous along the BEC line, jumping from positive values outside the BEC phase to negative values in the phase with a BC.
The above observations are valid for both scenarios.
In Scenario~II~($a > 0$) $S\sigma$ shows singular behavior at the CP.
The skewness can reach both $-\infty$ and  $+\infty$ at the CP depending on the path of approach.
When crossing the FOPT in Scenario II $S\sigma$ undergoes a jump discontinuity.

\paragraph*{Kurtosis.}
$\kappa\sigma^2$  presented in ~Fig.~\ref{fig-skewness}.
In both the scenarios it also always attains positive values everywhere on the phase diagram.
Kurtosis can strongly deviate from the baseline  $\kappa\sigma^2 = 1$ of an ideal Boltzmann gas.
This is due to the presence of interactions and Bose statistics.
The largest values of the kurtosis are generally obtained in the vicinity of the BEC line.
The kurtosis exhibits a non-monotonic behavior as a function of $\mu_I$ at both the BEC-line and the FOPT-line, where it jumps down as $\mu_I$ is increased.
$\kappa \sigma^2$ is an increasing function of $\mu_I$ elsewhere on the phase diagram.
The values of $\kappa\sigma^2$ remain large even far away from the CP and the Bose condensation boundary.
This is due to its large
sensitivity to interactions in the system.
$\kappa\sigma^2$ diverges at the CP.
The model does not predict negative values of $\kappa \sigma^2$ anywhere on the phase diagram.
This is in contrast to the universal behavior of fluctuations in the Ising model~\cite{Stephanov:2011pb,Bzdak:2016sxg}, as well as various model calculations~\cite{Vovchenko:2015pya,Vovchenko:2015uda,Chen:2015dra,Mukherjee:2016nhb,Motornenko:2019arp}, where negative values of $\kappa \sigma^2$ are observed in the so-called analytic crossover region above the critical temperature.
In the present work the negative values of $\kappa \sigma^2$ are not observed because of the Bose-Einstein condensation.
The BEC-line, which itself corresponds to a phase transition of a higher order, crosses the CP, thus no region in the vicinity of the CP can be identified as an analytic crossover.

We would like note that $\kappa\sigma^2$ exhibits a singular behavior also in Scenario I~($a=0$), at a  point ($T=0,\mu_I=m_{\pi})$~[see Fig.~\ref{fig-skewness} (c)].
Approaching this point along the BEC line one finds $\kappa\sigma^2 \rightarrow \infty$ at $T\rightarrow 0$. 

In the present work we do not discuss the behavior of fluctuations inside the mixed phase of the FOPT. 
These fluctuations can be addressed using the method developed in Ref.~\cite{Poberezhnyuk:2020cen} and will be the subject of a future study.

\section{Summary}
\label{sec-conc}

We studied thermodynamic properties of interacting pion matter in the framework of a mean-field model with a $\phi^4$-$\phi^6$ type Lagrangian.
The phase structure has been studied at non-zero isospin chemical potential $\mu_I$ that corresponds to the conserved 3rd component of isospin. 
Parameters of the repulsive and attractive interactions were fixed using 
lattice QCD data 
on the isospin density 
as the function of the chemical potential $\mu_I$ at zero temperature. 
The lattice data 
can be reasonably fitted within the two qualitatively different scenarios:
Scenario I with only repulsive interactions, and Scenario II with both repulsive and attractive interactions.
In both scenarios a phase with a Bose condensate of $\pi^+$ pions was found to occur at sufficiently large $\mu_I$, the transition between ordinary pion matter and matter with a BC taking place along the so-called BEC lines.
The presence of the attractive interactions in Scenario II leads,  in addition to the BEC, also to a first order liquid-gas phase transition of pions with a CP at $T_c\approx 0.369~m_\pi$ and $\mu_c\approx 0.991~m_\pi$.
A notable qualitative feature of the model, present for a broad range of values of parameters $a$ and $b$. is the fact that the BEC line merges with the FOPT line at the CP.
The system is characterized by two order parameters: (i) the difference $n_l - n_g >0 $ between the liquid and gas phase densities the corresponds to the liquid-gas transition; (ii) the density $n_+^{\rm bc}$ of the Bose condensed $\pi^+$ pions that characterizes the BEC transition.
This makes the model qualitatively different from the usual systems with a CP and FOPT where only a single order parameter is present.

The susceptibilities of isospin charge fluctuations up to the 4-th order
studied in the paper can serve as a robust observable to distinguish between the two  different scenarios. 
In the both scenarios, the scaled variance $\omega=\chi_2/\chi_1$,  skewness $S\sigma=\chi_3/\chi_2$, and kurtosis $\kappa\sigma^2=\chi_4/\chi_2$ 
remain finite on the BEC line. This happens due to the repulsive interactions 
in the pion system, in contrast to the ideal pion gas where these measures become infinite on the BEC line. 
All three fluctuation measures 
demonstrate
anomalous properties approaching the CP: $\omega\rightarrow \infty$, $\kappa\sigma^2\rightarrow \infty$, and $S\sigma$ can reach both
$+\infty$ and $-\infty$ depending on a  path to the CP. 
Note a significance of the higher order susceptibilities (e.g., skewness and kurtosis fluctuation measures) which are highly sensitive to a presence of the CP.
In the scenario I the CP is absent. In this case the anomalous fluctuations take place in the point $T=0$ and $\mu_I=m_\pi$.   
Approaching this point: $\omega$ and $\kappa\sigma^2$ can reach any value from 0 to $\infty$,  and $S\sigma$ can reach any value between $-\infty$ and $+\infty$ depending on a  path to the $(\mu_I=m_\pi,T=0)$--point. 
The susceptibilities $\chi_i$ can be computed in lattice QCD which are free of sign problem at finite $\mu_I$.
Analysis of their behavior can be used to establish a point~(region) of anomalously large fluctuations.
Determining whether this point corresponds to zero or finite temperatures  will allow to distinguish Scenarios I~($T = 0$) and II~($T=T_c > 0$). Besides, there is a qualitative difference in a behavior of the scaled variance $\omega$ near $T=0$ in Scenario I and near $T=T_c$ in Scenario II:  $\omega \rightarrow \infty$ at $T\rightarrow T_c$ in Scenario II, and $\omega$ can reach any value from 0 to $\infty$ depending on the path of approaching to $T=0$ in Scenario I.

The results obtained in this paper can be used in systems where the pion densities are large and a BC of pions may occur.
This can happen, for example, in heavy-ion or proton-proton collisions where the pion condensation may occur as a chemical non-equilibrium effect.
Other possibilities include pion stars as well as the Early Universe which may have passed through a pion-condensed phase if the lepton flavor asymmetries during its evolution were large.

\begin{acknowledgments}
We are grateful to D.V. Anchishkin,  I.N. Mishustin, L.M. Satarov,  and  H. Stoecker for fruitful discussions. 
This work is 
supported by the Target Program of Fundamental Research
of the Department of Physics and Astronomy of the National
Academy of Sciences of Ukraine (N 0120U100857).
The work of O.S.St. was partially  supported by the National Research Foundation of Ukraine under Project No. 2020.02/0073. 
\end{acknowledgments}

\bibliography{references.bib}

\begin{thebibliography}{64}%
\makeatletter
\providecommand \@ifxundefined [1]{%
 \@ifx{#1\undefined}
}%
\providecommand \@ifnum [1]{%
 \ifnum #1\expandafter \@firstoftwo
 \else \expandafter \@secondoftwo
 \fi
}%
\providecommand \@ifx [1]{%
 \ifx #1\expandafter \@firstoftwo
 \else \expandafter \@secondoftwo
 \fi
}%
\providecommand \natexlab [1]{#1}%
\providecommand \enquote  [1]{``#1''}%
\providecommand \bibnamefont  [1]{#1}%
\providecommand \bibfnamefont [1]{#1}%
\providecommand \citenamefont [1]{#1}%
\providecommand \href@noop [0]{\@secondoftwo}%
\providecommand \href [0]{\begingroup \@sanitize@url \@href}%
\providecommand \@href[1]{\@@startlink{#1}\@@href}%
\providecommand \@@href[1]{\endgroup#1\@@endlink}%
\providecommand \@sanitize@url [0]{\catcode `\\12\catcode `\$12\catcode
  `\&12\catcode `\#12\catcode `\^12\catcode `\_12\catcode `\%12\relax}%
\providecommand \@@startlink[1]{}%
\providecommand \@@endlink[0]{}%
\providecommand \url  [0]{\begingroup\@sanitize@url \@url }%
\providecommand \@url [1]{\endgroup\@href {#1}{\urlprefix }}%
\providecommand \urlprefix  [0]{URL }%
\providecommand \Eprint [0]{\href }%
\providecommand \doibase [0]{http://dx.doi.org/}%
\providecommand \selectlanguage [0]{\@gobble}%
\providecommand \bibinfo  [0]{\@secondoftwo}%
\providecommand \bibfield  [0]{\@secondoftwo}%
\providecommand \translation [1]{[#1]}%
\providecommand \BibitemOpen [0]{}%
\providecommand \bibitemStop [0]{}%
\providecommand \bibitemNoStop [0]{.\EOS\space}%
\providecommand \EOS [0]{\spacefactor3000\relax}%
\providecommand \BibitemShut  [1]{\csname bibitem#1\endcsname}%
\let\auto@bib@innerbib\@empty
\bibitem [{\citenamefont {Bose}(1924)}]{Bose:1924mk}%
  \BibitemOpen
  \bibfield  {author} {\bibinfo {author} {\bibfnamefont {S.~N.}\ \bibnamefont
  {Bose}},\ }\href {\doibase 10.1007/BF01327326} {\bibfield  {journal}
  {\bibinfo  {journal} {Z. Phys.}\ }\textbf {\bibinfo {volume} {26}},\ \bibinfo
  {pages} {178} (\bibinfo {year} {1924})}\BibitemShut {NoStop}%
\bibitem [{\citenamefont {Einstein}(1925)}]{einstein1925stizunger}%
  \BibitemOpen
  \bibfield  {author} {\bibinfo {author} {\bibfnamefont {A.}~\bibnamefont
  {Einstein}},\ }\href@noop {} {\bibfield  {journal} {\bibinfo  {journal} {Kgl.
  Preuss. Akad. Wiss}\ }\textbf {\bibinfo {volume} {1}},\ \bibinfo {pages}
  {137} (\bibinfo {year} {1925})}\BibitemShut {NoStop}%
\bibitem [{\citenamefont {Anderson}\ \emph {et~al.}(1995)\citenamefont
  {Anderson}, \citenamefont {Ensher}, \citenamefont {Matthews}, \citenamefont
  {Wieman},\ and\ \citenamefont {Cornell}}]{Anderson198}%
  \BibitemOpen
  \bibfield  {author} {\bibinfo {author} {\bibfnamefont {M.~H.}\ \bibnamefont
  {Anderson}}, \bibinfo {author} {\bibfnamefont {J.~R.}\ \bibnamefont
  {Ensher}}, \bibinfo {author} {\bibfnamefont {M.~R.}\ \bibnamefont
  {Matthews}}, \bibinfo {author} {\bibfnamefont {C.~E.}\ \bibnamefont
  {Wieman}}, \ and\ \bibinfo {author} {\bibfnamefont {E.~A.}\ \bibnamefont
  {Cornell}},\ }\href {\doibase 10.1126/science.269.5221.198} {\bibfield
  {journal} {\bibinfo  {journal} {Science}\ }\textbf {\bibinfo {volume}
  {269}},\ \bibinfo {pages} {198} (\bibinfo {year} {1995})}\BibitemShut
  {NoStop}%
\bibitem [{\citenamefont {Bradley}\ \emph {et~al.}(1995)\citenamefont
  {Bradley}, \citenamefont {Sackett}, \citenamefont {Tollett},\ and\
  \citenamefont {Hulet}}]{PhysRevLett.75.1687}%
  \BibitemOpen
  \bibfield  {author} {\bibinfo {author} {\bibfnamefont {C.~C.}\ \bibnamefont
  {Bradley}}, \bibinfo {author} {\bibfnamefont {C.~A.}\ \bibnamefont
  {Sackett}}, \bibinfo {author} {\bibfnamefont {J.~J.}\ \bibnamefont
  {Tollett}}, \ and\ \bibinfo {author} {\bibfnamefont {R.~G.}\ \bibnamefont
  {Hulet}},\ }\href {\doibase 10.1103/PhysRevLett.75.1687} {\bibfield
  {journal} {\bibinfo  {journal} {Phys. Rev. Lett.}\ }\textbf {\bibinfo
  {volume} {75}},\ \bibinfo {pages} {1687} (\bibinfo {year}
  {1995})}\BibitemShut {NoStop}%
\bibitem [{\citenamefont {Davis}\ \emph {et~al.}(1995)\citenamefont {Davis},
  \citenamefont {Mewes}, \citenamefont {Andrews}, \citenamefont {van Druten},
  \citenamefont {Durfee}, \citenamefont {Kurn},\ and\ \citenamefont
  {Ketterle}}]{PhysRevLett.75.3969}%
  \BibitemOpen
  \bibfield  {author} {\bibinfo {author} {\bibfnamefont {K.~B.}\ \bibnamefont
  {Davis}}, \bibinfo {author} {\bibfnamefont {M.~O.}\ \bibnamefont {Mewes}},
  \bibinfo {author} {\bibfnamefont {M.~R.}\ \bibnamefont {Andrews}}, \bibinfo
  {author} {\bibfnamefont {N.~J.}\ \bibnamefont {van Druten}}, \bibinfo
  {author} {\bibfnamefont {D.~S.}\ \bibnamefont {Durfee}}, \bibinfo {author}
  {\bibfnamefont {D.~M.}\ \bibnamefont {Kurn}}, \ and\ \bibinfo {author}
  {\bibfnamefont {W.}~\bibnamefont {Ketterle}},\ }\href {\doibase
  10.1103/PhysRevLett.75.3969} {\bibfield  {journal} {\bibinfo  {journal}
  {Phys. Rev. Lett.}\ }\textbf {\bibinfo {volume} {75}},\ \bibinfo {pages}
  {3969} (\bibinfo {year} {1995})}\BibitemShut {NoStop}%
\bibitem [{\citenamefont {Dalfovo}\ \emph {et~al.}(1999)\citenamefont
  {Dalfovo}, \citenamefont {Giorgini}, \citenamefont {Pitaevskii},\ and\
  \citenamefont {Stringari}}]{RevModPhys.71.463}%
  \BibitemOpen
  \bibfield  {author} {\bibinfo {author} {\bibfnamefont {F.}~\bibnamefont
  {Dalfovo}}, \bibinfo {author} {\bibfnamefont {S.}~\bibnamefont {Giorgini}},
  \bibinfo {author} {\bibfnamefont {L.~P.}\ \bibnamefont {Pitaevskii}}, \ and\
  \bibinfo {author} {\bibfnamefont {S.}~\bibnamefont {Stringari}},\ }\href
  {\doibase 10.1103/RevModPhys.71.463} {\bibfield  {journal} {\bibinfo
  {journal} {Rev. Mod. Phys.}\ }\textbf {\bibinfo {volume} {71}},\ \bibinfo
  {pages} {463} (\bibinfo {year} {1999})}\BibitemShut {NoStop}%
\bibitem [{\citenamefont {Satarov}\ \emph {et~al.}(2017)\citenamefont
  {Satarov}, \citenamefont {Gorenstein}, \citenamefont {Motornenko},
  \citenamefont {Vovchenko}, \citenamefont {Mishustin},\ and\ \citenamefont
  {Stoecker}}]{Satarov:2017jtu}%
  \BibitemOpen
  \bibfield  {author} {\bibinfo {author} {\bibfnamefont {L.}~\bibnamefont
  {Satarov}}, \bibinfo {author} {\bibfnamefont {M.}~\bibnamefont {Gorenstein}},
  \bibinfo {author} {\bibfnamefont {A.}~\bibnamefont {Motornenko}}, \bibinfo
  {author} {\bibfnamefont {V.}~\bibnamefont {Vovchenko}}, \bibinfo {author}
  {\bibfnamefont {I.}~\bibnamefont {Mishustin}}, \ and\ \bibinfo {author}
  {\bibfnamefont {H.}~\bibnamefont {Stoecker}},\ }\href {\doibase
  10.1088/1361-6471/aa8c5d} {\bibfield  {journal} {\bibinfo  {journal} {J.
  Phys. G}\ }\textbf {\bibinfo {volume} {44}},\ \bibinfo {pages} {12} (\bibinfo
  {year} {2017})},\ \Eprint {http://arxiv.org/abs/1704.08039} {arXiv:1704.08039
  [nucl-th]} \BibitemShut {NoStop}%
\bibitem [{\citenamefont {Begun}\ and\ \citenamefont
  {Gorenstein}(2007)}]{Begun:2006gj}%
  \BibitemOpen
  \bibfield  {author} {\bibinfo {author} {\bibfnamefont {V.}~\bibnamefont
  {Begun}}\ and\ \bibinfo {author} {\bibfnamefont {M.~I.}\ \bibnamefont
  {Gorenstein}},\ }\href {\doibase 10.1016/j.physletb.2007.07.059} {\bibfield
  {journal} {\bibinfo  {journal} {Phys. Lett. B}\ }\textbf {\bibinfo {volume}
  {653}},\ \bibinfo {pages} {190} (\bibinfo {year} {2007})},\ \Eprint
  {http://arxiv.org/abs/hep-ph/0611043} {arXiv:hep-ph/0611043} \BibitemShut
  {NoStop}%
\bibitem [{\citenamefont {Begun}\ and\ \citenamefont
  {Gorenstein}(2008)}]{Begun:2008hq}%
  \BibitemOpen
  \bibfield  {author} {\bibinfo {author} {\bibfnamefont {V.}~\bibnamefont
  {Begun}}\ and\ \bibinfo {author} {\bibfnamefont {M.}~\bibnamefont
  {Gorenstein}},\ }\href {\doibase 10.1103/PhysRevC.77.064903} {\bibfield
  {journal} {\bibinfo  {journal} {Phys. Rev. C}\ }\textbf {\bibinfo {volume}
  {77}},\ \bibinfo {pages} {064903} (\bibinfo {year} {2008})},\ \Eprint
  {http://arxiv.org/abs/0802.3349} {arXiv:0802.3349 [hep-ph]} \BibitemShut
  {NoStop}%
\bibitem [{\citenamefont {Strinati}\ \emph {et~al.}(2018)\citenamefont
  {Strinati}, \citenamefont {Pieri}, \citenamefont {Röpke}, \citenamefont
  {Schuck},\ and\ \citenamefont {Urban}}]{Strinati_2018}%
  \BibitemOpen
  \bibfield  {author} {\bibinfo {author} {\bibfnamefont {G.~C.}\ \bibnamefont
  {Strinati}}, \bibinfo {author} {\bibfnamefont {P.}~\bibnamefont {Pieri}},
  \bibinfo {author} {\bibfnamefont {G.}~\bibnamefont {Röpke}}, \bibinfo
  {author} {\bibfnamefont {P.}~\bibnamefont {Schuck}}, \ and\ \bibinfo {author}
  {\bibfnamefont {M.}~\bibnamefont {Urban}},\ }\href {\doibase
  https://doi.org/10.1016/j.physrep.2018.02.004} {\bibfield  {journal}
  {\bibinfo  {journal} {Physics Reports}\ }\textbf {\bibinfo {volume} {738}},\
  \bibinfo {pages} {1 } (\bibinfo {year} {2018})},\ \bibinfo {note} {the
  BCS–BEC crossover: From ultra-cold Fermi gases to nuclear
  systems}\BibitemShut {NoStop}%
\bibitem [{\citenamefont {Nozieres}\ and\ \citenamefont
  {Schmitt-Rink}(1985)}]{Nozieres:1985zz}%
  \BibitemOpen
  \bibfield  {author} {\bibinfo {author} {\bibfnamefont {P.}~\bibnamefont
  {Nozieres}}\ and\ \bibinfo {author} {\bibfnamefont {S.}~\bibnamefont
  {Schmitt-Rink}},\ }\href {\doibase 10.1007/BF00683774} {\bibfield  {journal}
  {\bibinfo  {journal} {J. Low. Temp. Phys.}\ }\textbf {\bibinfo {volume}
  {59}},\ \bibinfo {pages} {195} (\bibinfo {year} {1985})}\BibitemShut
  {NoStop}%
\bibitem [{\citenamefont {Funaki}\ \emph {et~al.}(2008)\citenamefont {Funaki},
  \citenamefont {Yamada}, \citenamefont {Horiuchi}, \citenamefont {R\"opke},
  \citenamefont {Schuck},\ and\ \citenamefont
  {Tohsaki}}]{PhysRevLett.101.082502}%
  \BibitemOpen
  \bibfield  {author} {\bibinfo {author} {\bibfnamefont {Y.}~\bibnamefont
  {Funaki}}, \bibinfo {author} {\bibfnamefont {T.}~\bibnamefont {Yamada}},
  \bibinfo {author} {\bibfnamefont {H.}~\bibnamefont {Horiuchi}}, \bibinfo
  {author} {\bibfnamefont {G.}~\bibnamefont {R\"opke}}, \bibinfo {author}
  {\bibfnamefont {P.}~\bibnamefont {Schuck}}, \ and\ \bibinfo {author}
  {\bibfnamefont {A.}~\bibnamefont {Tohsaki}},\ }\href {\doibase
  10.1103/PhysRevLett.101.082502} {\bibfield  {journal} {\bibinfo  {journal}
  {Phys. Rev. Lett.}\ }\textbf {\bibinfo {volume} {101}},\ \bibinfo {pages}
  {082502} (\bibinfo {year} {2008})}\BibitemShut {NoStop}%
\bibitem [{\citenamefont {Chavanis}\ and\ \citenamefont
  {Harko}(2012)}]{Chavanis:2011cz}%
  \BibitemOpen
  \bibfield  {author} {\bibinfo {author} {\bibfnamefont {P.-H.}\ \bibnamefont
  {Chavanis}}\ and\ \bibinfo {author} {\bibfnamefont {T.}~\bibnamefont
  {Harko}},\ }\href {\doibase 10.1103/PhysRevD.86.064011} {\bibfield  {journal}
  {\bibinfo  {journal} {Phys. Rev.}\ }\textbf {\bibinfo {volume} {D86}},\
  \bibinfo {pages} {064011} (\bibinfo {year} {2012})},\ \Eprint
  {http://arxiv.org/abs/1108.3986} {arXiv:1108.3986 [astro-ph.SR]} \BibitemShut
  {NoStop}%
\bibitem [{\citenamefont {Mishustin}\ \emph {et~al.}(2019)\citenamefont
  {Mishustin}, \citenamefont {Anchishkin}, \citenamefont {Satarov},
  \citenamefont {Stashko},\ and\ \citenamefont {Stoecker}}]{Mishustin:2019otg}%
  \BibitemOpen
  \bibfield  {author} {\bibinfo {author} {\bibfnamefont {I.}~\bibnamefont
  {Mishustin}}, \bibinfo {author} {\bibfnamefont {D.}~\bibnamefont
  {Anchishkin}}, \bibinfo {author} {\bibfnamefont {L.}~\bibnamefont {Satarov}},
  \bibinfo {author} {\bibfnamefont {O.}~\bibnamefont {Stashko}}, \ and\
  \bibinfo {author} {\bibfnamefont {H.}~\bibnamefont {Stoecker}},\ }\href
  {\doibase 10.1103/PhysRevC.100.022201} {\bibfield  {journal} {\bibinfo
  {journal} {Phys. Rev. C}\ }\textbf {\bibinfo {volume} {100}},\ \bibinfo
  {pages} {022201} (\bibinfo {year} {2019})},\ \Eprint
  {http://arxiv.org/abs/1905.09567} {arXiv:1905.09567 [nucl-th]} \BibitemShut
  {NoStop}%
\bibitem [{\citenamefont {Padilla}\ \emph {et~al.}(2019)\citenamefont
  {Padilla}, \citenamefont {Vázquez}, \citenamefont {Matos},\ and\
  \citenamefont {Germán}}]{Padilla_2019}%
  \BibitemOpen
  \bibfield  {author} {\bibinfo {author} {\bibfnamefont {L.~E.}\ \bibnamefont
  {Padilla}}, \bibinfo {author} {\bibfnamefont {J.~A.}\ \bibnamefont
  {Vázquez}}, \bibinfo {author} {\bibfnamefont {T.}~\bibnamefont {Matos}}, \
  and\ \bibinfo {author} {\bibfnamefont {G.}~\bibnamefont {Germán}},\ }\href
  {\doibase 10.1088/1475-7516/2019/05/056} {\bibfield  {journal} {\bibinfo
  {journal} {Journal of Cosmology and Astroparticle Physics}\ }\textbf
  {\bibinfo {volume} {2019}},\ \bibinfo {pages} {056–056} (\bibinfo {year}
  {2019})}\BibitemShut {NoStop}%
\bibitem [{\citenamefont {Kapusta}\ and\ \citenamefont
  {Gale}(2006)}]{kapusta_gale_2006}%
  \BibitemOpen
  \bibfield  {author} {\bibinfo {author} {\bibfnamefont {J.~I.}\ \bibnamefont
  {Kapusta}}\ and\ \bibinfo {author} {\bibfnamefont {C.}~\bibnamefont {Gale}},\
  }\href {\doibase 10.1017/CBO9780511535130} {\emph {\bibinfo {title}
  {Finite-Temperature Field Theory: Principles and Applications}}},\ \bibinfo
  {edition} {2nd}\ ed.,\ Cambridge Monographs on Mathematical Physics\
  (\bibinfo  {publisher} {Cambridge University Press},\ \bibinfo {year}
  {2006})\BibitemShut {NoStop}%
\bibitem [{\citenamefont {Andersen}(2004)}]{Andersen_2004}%
  \BibitemOpen
  \bibfield  {author} {\bibinfo {author} {\bibfnamefont {J.~O.}\ \bibnamefont
  {Andersen}},\ }\href {\doibase 10.1103/RevModPhys.76.599} {\bibfield
  {journal} {\bibinfo  {journal} {Rev. Mod. Phys.}\ }\textbf {\bibinfo {volume}
  {76}},\ \bibinfo {pages} {599} (\bibinfo {year} {2004})},\ \Eprint
  {http://arxiv.org/abs/cond-mat/0305138} {arXiv:cond-mat/0305138 [cond-mat]}
  \BibitemShut {NoStop}%
\bibitem [{\citenamefont {Griffin}\ \emph {et~al.}(1996)\citenamefont
  {Griffin}, \citenamefont {Snoke},\ and\ \citenamefont
  {Stringari}}]{griffin1996bose}%
  \BibitemOpen
  \bibfield  {author} {\bibinfo {author} {\bibfnamefont {A.}~\bibnamefont
  {Griffin}}, \bibinfo {author} {\bibfnamefont {D.~W.}\ \bibnamefont {Snoke}},
  \ and\ \bibinfo {author} {\bibfnamefont {S.}~\bibnamefont {Stringari}},\
  }\href {\doibase 10.1017/CBO9780511524240} {\emph {\bibinfo {title}
  {Bose-einstein condensation}}}\ (\bibinfo  {publisher} {Cambridge University
  Press},\ \bibinfo {year} {1996})\BibitemShut {NoStop}%
\bibitem [{\citenamefont {Watabe}\ and\ \citenamefont
  {Ohashi}(2013)}]{PhysRevA.88.053633}%
  \BibitemOpen
  \bibfield  {author} {\bibinfo {author} {\bibfnamefont {S.}~\bibnamefont
  {Watabe}}\ and\ \bibinfo {author} {\bibfnamefont {Y.}~\bibnamefont
  {Ohashi}},\ }\href {\doibase 10.1103/PhysRevA.88.053633} {\bibfield
  {journal} {\bibinfo  {journal} {Phys. Rev. A}\ }\textbf {\bibinfo {volume}
  {88}},\ \bibinfo {pages} {053633} (\bibinfo {year} {2013})}\BibitemShut
  {NoStop}%
\bibitem [{\citenamefont {Watabe}(2019)}]{Watabe_2019}%
  \BibitemOpen
  \bibfield  {author} {\bibinfo {author} {\bibfnamefont {S.}~\bibnamefont
  {Watabe}},\ }\href {\doibase 10.12693/aphyspola.135.1222} {\bibfield
  {journal} {\bibinfo  {journal} {Acta Phys. Polon. A}\ }\textbf {\bibinfo
  {volume} {135}},\ \bibinfo {pages} {1222–1230} (\bibinfo {year}
  {2019})}\BibitemShut {NoStop}%
\bibitem [{\citenamefont {Baym}\ \emph {et~al.}(1999)\citenamefont {Baym},
  \citenamefont {Blaizot}, \citenamefont {Holzmann}, \citenamefont {Lalo\"e},\
  and\ \citenamefont {Vautherin}}]{PhysRevLett.83.1703}%
  \BibitemOpen
  \bibfield  {author} {\bibinfo {author} {\bibfnamefont {G.}~\bibnamefont
  {Baym}}, \bibinfo {author} {\bibfnamefont {J.-P.}\ \bibnamefont {Blaizot}},
  \bibinfo {author} {\bibfnamefont {M.}~\bibnamefont {Holzmann}}, \bibinfo
  {author} {\bibfnamefont {F.}~\bibnamefont {Lalo\"e}}, \ and\ \bibinfo
  {author} {\bibfnamefont {D.}~\bibnamefont {Vautherin}},\ }\href {\doibase
  10.1103/PhysRevLett.83.1703} {\bibfield  {journal} {\bibinfo  {journal}
  {Phys. Rev. Lett.}\ }\textbf {\bibinfo {volume} {83}},\ \bibinfo {pages}
  {1703} (\bibinfo {year} {1999})}\BibitemShut {NoStop}%
\bibitem [{\citenamefont {Baym}\ \emph {et~al.}(2000)\citenamefont {Baym},
  \citenamefont {Blaizot},\ and\ \citenamefont {Zinn-Justin}}]{Baym_2000}%
  \BibitemOpen
  \bibfield  {author} {\bibinfo {author} {\bibfnamefont {G.}~\bibnamefont
  {Baym}}, \bibinfo {author} {\bibfnamefont {J.-P.}\ \bibnamefont {Blaizot}}, \
  and\ \bibinfo {author} {\bibfnamefont {J.}~\bibnamefont {Zinn-Justin}},\
  }\href {\doibase 10.1209/epl/i2000-00130-3} {\bibfield  {journal} {\bibinfo
  {journal} {EPL}\ }\textbf {\bibinfo {volume} {49}},\ \bibinfo {pages} {150}
  (\bibinfo {year} {2000})}\BibitemShut {NoStop}%
\bibitem [{\citenamefont {Holzmann}\ and\ \citenamefont
  {Krauth}(1999)}]{Holzmann_1999}%
  \BibitemOpen
  \bibfield  {author} {\bibinfo {author} {\bibfnamefont {M.}~\bibnamefont
  {Holzmann}}\ and\ \bibinfo {author} {\bibfnamefont {W.}~\bibnamefont
  {Krauth}},\ }\href {\doibase 10.1103/PhysRevLett.83.2687} {\bibfield
  {journal} {\bibinfo  {journal} {Phys. Rev. Lett.}\ }\textbf {\bibinfo
  {volume} {83}},\ \bibinfo {pages} {2687} (\bibinfo {year} {1999})},\ \Eprint
  {http://arxiv.org/abs/cond-mat/9905198} {arXiv:cond-mat/9905198
  [cond-mat.stat-mech]} \BibitemShut {NoStop}%
\bibitem [{\citenamefont {Holzmann}\ \emph {et~al.}(2001)\citenamefont
  {Holzmann}, \citenamefont {Baym}, \citenamefont {Blaizot},\ and\
  \citenamefont {Laloe}}]{Holzmann_2001}%
  \BibitemOpen
  \bibfield  {author} {\bibinfo {author} {\bibfnamefont {M.}~\bibnamefont
  {Holzmann}}, \bibinfo {author} {\bibfnamefont {G.}~\bibnamefont {Baym}},
  \bibinfo {author} {\bibfnamefont {J.-P.}\ \bibnamefont {Blaizot}}, \ and\
  \bibinfo {author} {\bibfnamefont {F.}~\bibnamefont {Laloe}},\ }\href
  {\doibase 10.1103/PhysRevLett.87.120403} {\bibfield  {journal} {\bibinfo
  {journal} {Phys. Rev. Lett.}\ }\textbf {\bibinfo {volume} {87}},\ \bibinfo
  {pages} {120403} (\bibinfo {year} {2001})},\ \Eprint
  {http://arxiv.org/abs/cond-mat/0103595} {arXiv:cond-mat/0103595
  [cond-mat.stat-mech]} \BibitemShut {NoStop}%
\bibitem [{\citenamefont {Huang}(1999)}]{PhysRevLett.83.3770}%
  \BibitemOpen
  \bibfield  {author} {\bibinfo {author} {\bibfnamefont {K.}~\bibnamefont
  {Huang}},\ }\href {\doibase 10.1103/PhysRevLett.83.3770} {\bibfield
  {journal} {\bibinfo  {journal} {Phys. Rev. Lett.}\ }\textbf {\bibinfo
  {volume} {83}},\ \bibinfo {pages} {3770} (\bibinfo {year}
  {1999})}\BibitemShut {NoStop}%
\bibitem [{\citenamefont {Son}\ and\ \citenamefont
  {Stephanov}(2001)}]{Son:2000xc}%
  \BibitemOpen
  \bibfield  {author} {\bibinfo {author} {\bibfnamefont {D.}~\bibnamefont
  {Son}}\ and\ \bibinfo {author} {\bibfnamefont {M.~A.}\ \bibnamefont
  {Stephanov}},\ }\href {\doibase 10.1103/PhysRevLett.86.592} {\bibfield
  {journal} {\bibinfo  {journal} {Phys. Rev. Lett.}\ }\textbf {\bibinfo
  {volume} {86}},\ \bibinfo {pages} {592} (\bibinfo {year} {2001})},\ \Eprint
  {http://arxiv.org/abs/hep-ph/0005225} {arXiv:hep-ph/0005225} \BibitemShut
  {NoStop}%
\bibitem [{\citenamefont {Abuki}\ \emph {et~al.}(2009)\citenamefont {Abuki},
  \citenamefont {Brauner},\ and\ \citenamefont {Warringa}}]{Abuki:2009hx}%
  \BibitemOpen
  \bibfield  {author} {\bibinfo {author} {\bibfnamefont {H.}~\bibnamefont
  {Abuki}}, \bibinfo {author} {\bibfnamefont {T.}~\bibnamefont {Brauner}}, \
  and\ \bibinfo {author} {\bibfnamefont {H.~J.}\ \bibnamefont {Warringa}},\
  }\href {\doibase 10.1140/epjc/s10052-009-1121-0} {\bibfield  {journal}
  {\bibinfo  {journal} {Eur. Phys. J. C}\ }\textbf {\bibinfo {volume} {64}},\
  \bibinfo {pages} {123} (\bibinfo {year} {2009})},\ \Eprint
  {http://arxiv.org/abs/0901.2477} {arXiv:0901.2477 [hep-ph]} \BibitemShut
  {NoStop}%
\bibitem [{\citenamefont {Brandt}\ \emph
  {et~al.}(2018{\natexlab{a}})\citenamefont {Brandt}, \citenamefont {Endrodi},\
  and\ \citenamefont {Schmalzbauer}}]{Brandt:2017oyy}%
  \BibitemOpen
  \bibfield  {author} {\bibinfo {author} {\bibfnamefont {B.}~\bibnamefont
  {Brandt}}, \bibinfo {author} {\bibfnamefont {G.}~\bibnamefont {Endrodi}}, \
  and\ \bibinfo {author} {\bibfnamefont {S.}~\bibnamefont {Schmalzbauer}},\
  }\href {\doibase 10.1103/PhysRevD.97.054514} {\bibfield  {journal} {\bibinfo
  {journal} {Phys. Rev. D}\ }\textbf {\bibinfo {volume} {97}},\ \bibinfo
  {pages} {054514} (\bibinfo {year} {2018}{\natexlab{a}})},\ \Eprint
  {http://arxiv.org/abs/1712.08190} {arXiv:1712.08190 [hep-lat]} \BibitemShut
  {NoStop}%
\bibitem [{\citenamefont {Brandt}\ \emph
  {et~al.}(2018{\natexlab{b}})\citenamefont {Brandt}, \citenamefont {Endrodi},
  \citenamefont {Fraga}, \citenamefont {Hippert}, \citenamefont
  {Schaffner-Bielich},\ and\ \citenamefont {Schmalzbauer}}]{Brandt:2018bwq}%
  \BibitemOpen
  \bibfield  {author} {\bibinfo {author} {\bibfnamefont {B.~B.}\ \bibnamefont
  {Brandt}}, \bibinfo {author} {\bibfnamefont {G.}~\bibnamefont {Endrodi}},
  \bibinfo {author} {\bibfnamefont {E.~S.}\ \bibnamefont {Fraga}}, \bibinfo
  {author} {\bibfnamefont {M.}~\bibnamefont {Hippert}}, \bibinfo {author}
  {\bibfnamefont {J.}~\bibnamefont {Schaffner-Bielich}}, \ and\ \bibinfo
  {author} {\bibfnamefont {S.}~\bibnamefont {Schmalzbauer}},\ }\href {\doibase
  10.1103/PhysRevD.98.094510} {\bibfield  {journal} {\bibinfo  {journal} {Phys.
  Rev. D}\ }\textbf {\bibinfo {volume} {98}},\ \bibinfo {pages} {094510}
  (\bibinfo {year} {2018}{\natexlab{b}})},\ \Eprint
  {http://arxiv.org/abs/1802.06685} {arXiv:1802.06685 [hep-ph]} \BibitemShut
  {NoStop}%
\bibitem [{\citenamefont {Vovchenko}\ \emph {et~al.}(2021)\citenamefont
  {Vovchenko}, \citenamefont {Brandt}, \citenamefont {Cuteri}, \citenamefont
  {Endr\H{o}di}, \citenamefont {Hajkarim},\ and\ \citenamefont
  {Schaffner-Bielich}}]{Vovchenko:2020crk}%
  \BibitemOpen
  \bibfield  {author} {\bibinfo {author} {\bibfnamefont {V.}~\bibnamefont
  {Vovchenko}}, \bibinfo {author} {\bibfnamefont {B.~B.}\ \bibnamefont
  {Brandt}}, \bibinfo {author} {\bibfnamefont {F.}~\bibnamefont {Cuteri}},
  \bibinfo {author} {\bibfnamefont {G.}~\bibnamefont {Endr\H{o}di}}, \bibinfo
  {author} {\bibfnamefont {F.}~\bibnamefont {Hajkarim}}, \ and\ \bibinfo
  {author} {\bibfnamefont {J.}~\bibnamefont {Schaffner-Bielich}},\ }\href
  {\doibase 10.1103/PhysRevLett.126.012701} {\bibfield  {journal} {\bibinfo
  {journal} {Phys. Rev. Lett.}\ }\textbf {\bibinfo {volume} {126}},\ \bibinfo
  {pages} {012701} (\bibinfo {year} {2021})},\ \Eprint
  {http://arxiv.org/abs/2009.02309} {arXiv:2009.02309 [hep-ph]} \BibitemShut
  {NoStop}%
\bibitem [{\citenamefont {Mannarelli}(2019)}]{Mannarelli_2019}%
  \BibitemOpen
  \bibfield  {author} {\bibinfo {author} {\bibfnamefont {M.}~\bibnamefont
  {Mannarelli}},\ }\href {\doibase 10.3390/particles2030025} {\bibfield
  {journal} {\bibinfo  {journal} {Particles}\ }\textbf {\bibinfo {volume}
  {2}},\ \bibinfo {pages} {411–443} (\bibinfo {year} {2019})}\BibitemShut
  {NoStop}%
\bibitem [{\citenamefont {Andersen}\ and\ \citenamefont
  {Kneschke}(2018)}]{andersen2018boseeinstein}%
  \BibitemOpen
  \bibfield  {author} {\bibinfo {author} {\bibfnamefont {J.~O.}\ \bibnamefont
  {Andersen}}\ and\ \bibinfo {author} {\bibfnamefont {P.}~\bibnamefont
  {Kneschke}},\ }\href@noop {} {\enquote {\bibinfo {title} {Bose-einstein
  condensation and pion stars},}\ } (\bibinfo {year} {2018}),\ \Eprint
  {http://arxiv.org/abs/1807.08951} {arXiv:1807.08951 [hep-ph]} \BibitemShut
  {NoStop}%
\bibitem [{\citenamefont {Begun}\ and\ \citenamefont
  {Florkowski}(2015)}]{Begun:2015ifa}%
  \BibitemOpen
  \bibfield  {author} {\bibinfo {author} {\bibfnamefont {V.}~\bibnamefont
  {Begun}}\ and\ \bibinfo {author} {\bibfnamefont {W.}~\bibnamefont
  {Florkowski}},\ }\href {\doibase 10.1103/PhysRevC.91.054909} {\bibfield
  {journal} {\bibinfo  {journal} {Phys. Rev. C}\ }\textbf {\bibinfo {volume}
  {91}},\ \bibinfo {pages} {054909} (\bibinfo {year} {2015})},\ \Eprint
  {http://arxiv.org/abs/1503.04040} {arXiv:1503.04040 [nucl-th]} \BibitemShut
  {NoStop}%
\bibitem [{\citenamefont {Schunck}\ and\ \citenamefont
  {Mielke}(2003)}]{Schunck_2003}%
  \BibitemOpen
  \bibfield  {author} {\bibinfo {author} {\bibfnamefont {F.~E.}\ \bibnamefont
  {Schunck}}\ and\ \bibinfo {author} {\bibfnamefont {E.~W.}\ \bibnamefont
  {Mielke}},\ }\href {\doibase 10.1088/0264-9381/20/20/201} {\bibfield
  {journal} {\bibinfo  {journal} {Classical and Quantum Gravity}\ }\textbf
  {\bibinfo {volume} {20}},\ \bibinfo {pages} {R301–R356} (\bibinfo {year}
  {2003})}\BibitemShut {NoStop}%
\bibitem [{\citenamefont {{Liebling}}\ and\ \citenamefont
  {{Palenzuela}}(2017)}]{2017Liebling}%
  \BibitemOpen
  \bibfield  {author} {\bibinfo {author} {\bibfnamefont {S.~L.}\ \bibnamefont
  {{Liebling}}}\ and\ \bibinfo {author} {\bibfnamefont {C.}~\bibnamefont
  {{Palenzuela}}},\ }\href {\doibase 10.1007/s41114-017-0007-y} {\bibfield
  {journal} {\bibinfo  {journal} {Living Reviews in Relativity}\ }\textbf
  {\bibinfo {volume} {20}},\ \bibinfo {eid} {5} (\bibinfo {year}
  {2017})}\BibitemShut {NoStop}%
\bibitem [{\citenamefont {Braaten}\ \emph {et~al.}(2016)\citenamefont
  {Braaten}, \citenamefont {Mohapatra},\ and\ \citenamefont
  {Zhang}}]{Braaten_2016}%
  \BibitemOpen
  \bibfield  {author} {\bibinfo {author} {\bibfnamefont {E.}~\bibnamefont
  {Braaten}}, \bibinfo {author} {\bibfnamefont {A.}~\bibnamefont {Mohapatra}},
  \ and\ \bibinfo {author} {\bibfnamefont {H.}~\bibnamefont {Zhang}},\ }\href
  {\doibase 10.1103/physrevlett.117.121801} {\bibfield  {journal} {\bibinfo
  {journal} {Physical Review Letters}\ }\textbf {\bibinfo {volume} {117}}
  (\bibinfo {year} {2016}),\ 10.1103/physrevlett.117.121801}\BibitemShut
  {NoStop}%
\bibitem [{\citenamefont {Suárez}\ \emph {et~al.}(2013)\citenamefont
  {Suárez}, \citenamefont {Robles},\ and\ \citenamefont
  {Matos}}]{Su_rez_2013}%
  \BibitemOpen
  \bibfield  {author} {\bibinfo {author} {\bibfnamefont {A.}~\bibnamefont
  {Suárez}}, \bibinfo {author} {\bibfnamefont {V.~H.}\ \bibnamefont {Robles}},
  \ and\ \bibinfo {author} {\bibfnamefont {T.}~\bibnamefont {Matos}},\ }\href
  {\doibase 10.1007/978-3-319-02063-1_9} {\bibfield  {journal} {\bibinfo
  {journal} {Accelerated Cosmic Expansion}\ ,\ \bibinfo {pages} {107–142}}
  (\bibinfo {year} {2013})}\BibitemShut {NoStop}%
\bibitem [{\citenamefont {Bernal}\ \emph {et~al.}(2017)\citenamefont {Bernal},
  \citenamefont {Robles},\ and\ \citenamefont {Matos}}]{Bernal_2017}%
  \BibitemOpen
  \bibfield  {author} {\bibinfo {author} {\bibfnamefont {T.}~\bibnamefont
  {Bernal}}, \bibinfo {author} {\bibfnamefont {V.~H.}\ \bibnamefont {Robles}},
  \ and\ \bibinfo {author} {\bibfnamefont {T.}~\bibnamefont {Matos}},\ }\href
  {\doibase 10.1093/mnras/stx651} {\bibfield  {journal} {\bibinfo  {journal}
  {Monthly Notices of the Royal Astronomical Society}\ }\textbf {\bibinfo
  {volume} {468}},\ \bibinfo {pages} {3135–3149} (\bibinfo {year}
  {2017})}\BibitemShut {NoStop}%
\bibitem [{\citenamefont {Visinelli}(2016)}]{Visinelli_2016}%
  \BibitemOpen
  \bibfield  {author} {\bibinfo {author} {\bibfnamefont {L.}~\bibnamefont
  {Visinelli}},\ }\href {\doibase 10.1088/1475-7516/2016/07/009} {\bibfield
  {journal} {\bibinfo  {journal} {Journal of Cosmology and Astroparticle
  Physics}\ }\textbf {\bibinfo {volume} {2016}},\ \bibinfo {pages} {009–009}
  (\bibinfo {year} {2016})}\BibitemShut {NoStop}%
\bibitem [{\citenamefont {HajiSadeghi}\ \emph {et~al.}(2019)\citenamefont
  {HajiSadeghi}, \citenamefont {Smolenski},\ and\ \citenamefont
  {Wudka}}]{HajiSadeghi_2019}%
  \BibitemOpen
  \bibfield  {author} {\bibinfo {author} {\bibfnamefont {S.}~\bibnamefont
  {HajiSadeghi}}, \bibinfo {author} {\bibfnamefont {S.}~\bibnamefont
  {Smolenski}}, \ and\ \bibinfo {author} {\bibfnamefont {J.}~\bibnamefont
  {Wudka}},\ }\href {\doibase 10.1103/physrevd.99.023514} {\bibfield  {journal}
  {\bibinfo  {journal} {Physical Review D}\ }\textbf {\bibinfo {volume} {99}}
  (\bibinfo {year} {2019}),\ 10.1103/physrevd.99.023514}\BibitemShut {NoStop}%
\bibitem [{\citenamefont {{Barranco}}\ and\ \citenamefont
  {{Bernal}}(2011)}]{2011PhRvD..83d3525B}%
  \BibitemOpen
  \bibfield  {author} {\bibinfo {author} {\bibfnamefont {J.}~\bibnamefont
  {{Barranco}}}\ and\ \bibinfo {author} {\bibfnamefont {A.}~\bibnamefont
  {{Bernal}}},\ }\href {\doibase 10.1103/PhysRevD.83.043525} {\bibfield
  {journal} {\bibinfo  {journal} {\prd}\ }\textbf {\bibinfo {volume} {83}},\
  \bibinfo {eid} {043525} (\bibinfo {year} {2011})},\ \Eprint
  {http://arxiv.org/abs/1001.1769} {arXiv:1001.1769 [astro-ph.CO]} \BibitemShut
  {NoStop}%
\bibitem [{\citenamefont {Gavrilik}\ \emph {et~al.}(2020)\citenamefont
  {Gavrilik}, \citenamefont {Khelashvili},\ and\ \citenamefont
  {Nazarenko}}]{Gavrilik_2020}%
  \BibitemOpen
  \bibfield  {author} {\bibinfo {author} {\bibfnamefont {A.}~\bibnamefont
  {Gavrilik}}, \bibinfo {author} {\bibfnamefont {M.}~\bibnamefont
  {Khelashvili}}, \ and\ \bibinfo {author} {\bibfnamefont {A.}~\bibnamefont
  {Nazarenko}},\ }\href {\doibase 10.1103/physrevd.102.083510} {\bibfield
  {journal} {\bibinfo  {journal} {Physical Review D}\ }\textbf {\bibinfo
  {volume} {102}} (\bibinfo {year} {2020}),\
  10.1103/physrevd.102.083510}\BibitemShut {NoStop}%
\bibitem [{\citenamefont {Adhikari}\ and\ \citenamefont
  {Andersen}(2020)}]{Adhikari:2019zaj}%
  \BibitemOpen
  \bibfield  {author} {\bibinfo {author} {\bibfnamefont {P.}~\bibnamefont
  {Adhikari}}\ and\ \bibinfo {author} {\bibfnamefont {J.~O.}\ \bibnamefont
  {Andersen}},\ }\href {\doibase 10.1016/j.physletb.2020.135352} {\bibfield
  {journal} {\bibinfo  {journal} {Phys. Lett. B}\ }\textbf {\bibinfo {volume}
  {804}},\ \bibinfo {pages} {135352} (\bibinfo {year} {2020})},\ \Eprint
  {http://arxiv.org/abs/1909.01131} {arXiv:1909.01131 [hep-ph]} \BibitemShut
  {NoStop}%
\bibitem [{\citenamefont {Adhikari}\ \emph {et~al.}(2020)\citenamefont
  {Adhikari}, \citenamefont {Andersen},\ and\ \citenamefont
  {Mojahed}}]{Adhikari:2020kdn}%
  \BibitemOpen
  \bibfield  {author} {\bibinfo {author} {\bibfnamefont {P.}~\bibnamefont
  {Adhikari}}, \bibinfo {author} {\bibfnamefont {J.~O.}\ \bibnamefont
  {Andersen}}, \ and\ \bibinfo {author} {\bibfnamefont {M.~A.}\ \bibnamefont
  {Mojahed}},\ }\href@noop {} {\  (\bibinfo {year} {2020})},\ \Eprint
  {http://arxiv.org/abs/2010.13655} {arXiv:2010.13655 [hep-ph]} \BibitemShut
  {NoStop}%
\bibitem [{\citenamefont {He}\ \emph {et~al.}(2005)\citenamefont {He},
  \citenamefont {Jin},\ and\ \citenamefont {Zhuang}}]{He:2005nk}%
  \BibitemOpen
  \bibfield  {author} {\bibinfo {author} {\bibfnamefont {L.-y.}\ \bibnamefont
  {He}}, \bibinfo {author} {\bibfnamefont {M.}~\bibnamefont {Jin}}, \ and\
  \bibinfo {author} {\bibfnamefont {P.-f.}\ \bibnamefont {Zhuang}},\ }\href
  {\doibase 10.1103/PhysRevD.71.116001} {\bibfield  {journal} {\bibinfo
  {journal} {Phys. Rev. D}\ }\textbf {\bibinfo {volume} {71}},\ \bibinfo
  {pages} {116001} (\bibinfo {year} {2005})},\ \Eprint
  {http://arxiv.org/abs/hep-ph/0503272} {arXiv:hep-ph/0503272} \BibitemShut
  {NoStop}%
\bibitem [{\citenamefont {Adhikari}\ \emph {et~al.}(2018)\citenamefont
  {Adhikari}, \citenamefont {Andersen},\ and\ \citenamefont
  {Kneschke}}]{Adhikari:2018cea}%
  \BibitemOpen
  \bibfield  {author} {\bibinfo {author} {\bibfnamefont {P.}~\bibnamefont
  {Adhikari}}, \bibinfo {author} {\bibfnamefont {J.~O.}\ \bibnamefont
  {Andersen}}, \ and\ \bibinfo {author} {\bibfnamefont {P.}~\bibnamefont
  {Kneschke}},\ }\href {\doibase 10.1103/PhysRevD.98.074016} {\bibfield
  {journal} {\bibinfo  {journal} {Phys. Rev. D}\ }\textbf {\bibinfo {volume}
  {98}},\ \bibinfo {pages} {074016} (\bibinfo {year} {2018})},\ \Eprint
  {http://arxiv.org/abs/1805.08599} {arXiv:1805.08599 [hep-ph]} \BibitemShut
  {NoStop}%
\bibitem [{\citenamefont {Folkestad}\ and\ \citenamefont
  {Andersen}(2019)}]{Folkestad:2018psc}%
  \BibitemOpen
  \bibfield  {author} {\bibinfo {author} {\bibfnamefont {A.}~\bibnamefont
  {Folkestad}}\ and\ \bibinfo {author} {\bibfnamefont {J.~O.}\ \bibnamefont
  {Andersen}},\ }\href {\doibase 10.1103/PhysRevD.99.054006} {\bibfield
  {journal} {\bibinfo  {journal} {Phys. Rev. D}\ }\textbf {\bibinfo {volume}
  {99}},\ \bibinfo {pages} {054006} (\bibinfo {year} {2019})},\ \Eprint
  {http://arxiv.org/abs/1810.10573} {arXiv:1810.10573 [hep-ph]} \BibitemShut
  {NoStop}%
\bibitem [{\citenamefont {Anchishkin}\ \emph {et~al.}(2019)\citenamefont
  {Anchishkin}, \citenamefont {Mishustin},\ and\ \citenamefont
  {Stoecker}}]{Anchishkin_2019}%
  \BibitemOpen
  \bibfield  {author} {\bibinfo {author} {\bibfnamefont {D.}~\bibnamefont
  {Anchishkin}}, \bibinfo {author} {\bibfnamefont {I.}~\bibnamefont
  {Mishustin}}, \ and\ \bibinfo {author} {\bibfnamefont {H.}~\bibnamefont
  {Stoecker}},\ }\href {\doibase 10.1088/1361-6471/aafea8} {\bibfield
  {journal} {\bibinfo  {journal} {J.\ Phys.\ G}\ }\textbf {\bibinfo {volume}
  {46}},\ \bibinfo {pages} {035002} (\bibinfo {year} {2019})},\ \Eprint
  {http://arxiv.org/abs/1806.10857} {arXiv:1806.10857 [nucl-th]} \BibitemShut
  {NoStop}%
\bibitem [{\citenamefont {Stashko}\ \emph {et~al.}(2020)\citenamefont
  {Stashko}, \citenamefont {Anchishkin}, \citenamefont {Savchuk},\ and\
  \citenamefont {Gorenstein}}]{stashko2020thermodynamic}%
  \BibitemOpen
  \bibfield  {author} {\bibinfo {author} {\bibfnamefont {O.~S.}\ \bibnamefont
  {Stashko}}, \bibinfo {author} {\bibfnamefont {D.~V.}\ \bibnamefont
  {Anchishkin}}, \bibinfo {author} {\bibfnamefont {O.~V.}\ \bibnamefont
  {Savchuk}}, \ and\ \bibinfo {author} {\bibfnamefont {M.~I.}\ \bibnamefont
  {Gorenstein}},\ }\href@noop {} {\enquote {\bibinfo {title} {Thermodynamic
  properties of interacting bosons with zero chemical potential},}\ } (\bibinfo
  {year} {2020}),\ \Eprint {http://arxiv.org/abs/2007.06321} {arXiv:2007.06321
  [hep-ph]} \BibitemShut {NoStop}%
\bibitem [{\citenamefont {Savchuk}\ \emph {et~al.}(2020)\citenamefont
  {Savchuk}, \citenamefont {Bondar}, \citenamefont {Stashko}, \citenamefont
  {Poberezhnyuk}, \citenamefont {Vovchenko}, \citenamefont {Gorenstein},\ and\
  \citenamefont {Stoecker}}]{Savchuk:2020yxc}%
  \BibitemOpen
  \bibfield  {author} {\bibinfo {author} {\bibfnamefont {O.}~\bibnamefont
  {Savchuk}}, \bibinfo {author} {\bibfnamefont {Y.}~\bibnamefont {Bondar}},
  \bibinfo {author} {\bibfnamefont {O.}~\bibnamefont {Stashko}}, \bibinfo
  {author} {\bibfnamefont {R.~V.}\ \bibnamefont {Poberezhnyuk}}, \bibinfo
  {author} {\bibfnamefont {V.}~\bibnamefont {Vovchenko}}, \bibinfo {author}
  {\bibfnamefont {M.~I.}\ \bibnamefont {Gorenstein}}, \ and\ \bibinfo {author}
  {\bibfnamefont {H.}~\bibnamefont {Stoecker}},\ }\href {\doibase
  10.1103/PhysRevC.102.035202} {\bibfield  {journal} {\bibinfo  {journal}
  {Phys. Rev. C}\ }\textbf {\bibinfo {volume} {102}},\ \bibinfo {pages}
  {035202} (\bibinfo {year} {2020})},\ \Eprint
  {http://arxiv.org/abs/2004.09004} {arXiv:2004.09004 [hep-ph]} \BibitemShut
  {NoStop}%
\bibitem [{\citenamefont {Satarov}\ \emph {et~al.}(2020)\citenamefont
  {Satarov}, \citenamefont {Poberezhnyuk}, \citenamefont {Mishustin},\ and\
  \citenamefont {Stoecker}}]{Satarov:2020loq}%
  \BibitemOpen
  \bibfield  {author} {\bibinfo {author} {\bibfnamefont {L.~M.}\ \bibnamefont
  {Satarov}}, \bibinfo {author} {\bibfnamefont {R.~V.}\ \bibnamefont
  {Poberezhnyuk}}, \bibinfo {author} {\bibfnamefont {I.~N.}\ \bibnamefont
  {Mishustin}}, \ and\ \bibinfo {author} {\bibfnamefont {H.}~\bibnamefont
  {Stoecker}},\ }\href@noop {} {\  (\bibinfo {year} {2020})},\ \Eprint
  {http://arxiv.org/abs/2009.13487} {arXiv:2009.13487 [nucl-th]} \BibitemShut
  {NoStop}%
\bibitem [{\citenamefont {Borsanyi}\ \emph {et~al.}(2014)\citenamefont
  {Borsanyi}, \citenamefont {Fodor}, \citenamefont {Hoelbling}, \citenamefont
  {Katz}, \citenamefont {Krieg},\ and\ \citenamefont
  {Szabo}}]{Borsanyi:2013bia}%
  \BibitemOpen
  \bibfield  {author} {\bibinfo {author} {\bibfnamefont {S.}~\bibnamefont
  {Borsanyi}}, \bibinfo {author} {\bibfnamefont {Z.}~\bibnamefont {Fodor}},
  \bibinfo {author} {\bibfnamefont {C.}~\bibnamefont {Hoelbling}}, \bibinfo
  {author} {\bibfnamefont {S.~D.}\ \bibnamefont {Katz}}, \bibinfo {author}
  {\bibfnamefont {S.}~\bibnamefont {Krieg}}, \ and\ \bibinfo {author}
  {\bibfnamefont {K.~K.}\ \bibnamefont {Szabo}},\ }\href {\doibase
  10.1016/j.physletb.2014.01.007} {\bibfield  {journal} {\bibinfo  {journal}
  {Phys. Lett. B}\ }\textbf {\bibinfo {volume} {730}},\ \bibinfo {pages} {99}
  (\bibinfo {year} {2014})},\ \Eprint {http://arxiv.org/abs/1309.5258}
  {arXiv:1309.5258 [hep-lat]} \BibitemShut {NoStop}%
\bibitem [{\citenamefont {Bazavov}\ \emph {et~al.}(2014)\citenamefont {Bazavov}
  \emph {et~al.}}]{Bazavov:2014pvz}%
  \BibitemOpen
  \bibfield  {author} {\bibinfo {author} {\bibfnamefont {A.}~\bibnamefont
  {Bazavov}} \emph {et~al.} (\bibinfo {collaboration} {HotQCD}),\ }\href
  {\doibase 10.1103/PhysRevD.90.094503} {\bibfield  {journal} {\bibinfo
  {journal} {Phys. Rev. D}\ }\textbf {\bibinfo {volume} {90}},\ \bibinfo
  {pages} {094503} (\bibinfo {year} {2014})},\ \Eprint
  {http://arxiv.org/abs/1407.6387} {arXiv:1407.6387 [hep-lat]} \BibitemShut
  {NoStop}%
\bibitem [{\citenamefont {Vovchenko}\ \emph
  {et~al.}(2015{\natexlab{a}})\citenamefont {Vovchenko}, \citenamefont
  {Anchishkin},\ and\ \citenamefont {Gorenstein}}]{Vovchenko:2014pka}%
  \BibitemOpen
  \bibfield  {author} {\bibinfo {author} {\bibfnamefont {V.}~\bibnamefont
  {Vovchenko}}, \bibinfo {author} {\bibfnamefont {D.~V.}\ \bibnamefont
  {Anchishkin}}, \ and\ \bibinfo {author} {\bibfnamefont {M.~I.}\ \bibnamefont
  {Gorenstein}},\ }\href {\doibase 10.1103/PhysRevC.91.024905} {\bibfield
  {journal} {\bibinfo  {journal} {Phys. Rev. C}\ }\textbf {\bibinfo {volume}
  {91}},\ \bibinfo {pages} {024905} (\bibinfo {year} {2015}{\natexlab{a}})},\
  \Eprint {http://arxiv.org/abs/1412.5478} {arXiv:1412.5478 [nucl-th]}
  \BibitemShut {NoStop}%
\bibitem [{\citenamefont {Karsch}\ and\ \citenamefont
  {Redlich}(2011)}]{Karsch:2010ck}%
  \BibitemOpen
  \bibfield  {author} {\bibinfo {author} {\bibfnamefont {F.}~\bibnamefont
  {Karsch}}\ and\ \bibinfo {author} {\bibfnamefont {K.}~\bibnamefont
  {Redlich}},\ }\href {\doibase 10.1016/j.physletb.2010.10.046} {\bibfield
  {journal} {\bibinfo  {journal} {Phys. Lett. B}\ }\textbf {\bibinfo {volume}
  {695}},\ \bibinfo {pages} {136} (\bibinfo {year} {2011})},\ \Eprint
  {http://arxiv.org/abs/1007.2581} {arXiv:1007.2581 [hep-ph]} \BibitemShut
  {NoStop}%
\bibitem [{\citenamefont {Ivanov}(2009)}]{PhysRevE.79.021116}%
  \BibitemOpen
  \bibfield  {author} {\bibinfo {author} {\bibfnamefont {I.~P.}\ \bibnamefont
  {Ivanov}},\ }\href {\doibase 10.1103/PhysRevE.79.021116} {\bibfield
  {journal} {\bibinfo  {journal} {Phys. Rev. E}\ }\textbf {\bibinfo {volume}
  {79}},\ \bibinfo {pages} {021116} (\bibinfo {year} {2009})}\BibitemShut
  {NoStop}%
\bibitem [{\citenamefont {Stephanov}(2011)}]{Stephanov:2011pb}%
  \BibitemOpen
  \bibfield  {author} {\bibinfo {author} {\bibfnamefont {M.~A.}\ \bibnamefont
  {Stephanov}},\ }\href {\doibase 10.1103/PhysRevLett.107.052301} {\bibfield
  {journal} {\bibinfo  {journal} {Phys. Rev. Lett.}\ }\textbf {\bibinfo
  {volume} {107}},\ \bibinfo {pages} {052301} (\bibinfo {year} {2011})},\
  \Eprint {http://arxiv.org/abs/1104.1627} {arXiv:1104.1627 [hep-ph]}
  \BibitemShut {NoStop}%
\bibitem [{\citenamefont {Bzdak}\ \emph {et~al.}(2017)\citenamefont {Bzdak},
  \citenamefont {Koch},\ and\ \citenamefont {Strodthoff}}]{Bzdak:2016sxg}%
  \BibitemOpen
  \bibfield  {author} {\bibinfo {author} {\bibfnamefont {A.}~\bibnamefont
  {Bzdak}}, \bibinfo {author} {\bibfnamefont {V.}~\bibnamefont {Koch}}, \ and\
  \bibinfo {author} {\bibfnamefont {N.}~\bibnamefont {Strodthoff}},\ }\href
  {\doibase 10.1103/PhysRevC.95.054906} {\bibfield  {journal} {\bibinfo
  {journal} {Phys. Rev. C}\ }\textbf {\bibinfo {volume} {95}},\ \bibinfo
  {pages} {054906} (\bibinfo {year} {2017})},\ \Eprint
  {http://arxiv.org/abs/1607.07375} {arXiv:1607.07375 [nucl-th]} \BibitemShut
  {NoStop}%
\bibitem [{\citenamefont {Vovchenko}\ \emph
  {et~al.}(2015{\natexlab{b}})\citenamefont {Vovchenko}, \citenamefont
  {Anchishkin}, \citenamefont {Gorenstein},\ and\ \citenamefont
  {Poberezhnyuk}}]{Vovchenko:2015pya}%
  \BibitemOpen
  \bibfield  {author} {\bibinfo {author} {\bibfnamefont {V.}~\bibnamefont
  {Vovchenko}}, \bibinfo {author} {\bibfnamefont {D.}~\bibnamefont
  {Anchishkin}}, \bibinfo {author} {\bibfnamefont {M.}~\bibnamefont
  {Gorenstein}}, \ and\ \bibinfo {author} {\bibfnamefont {R.}~\bibnamefont
  {Poberezhnyuk}},\ }\href {\doibase 10.1103/PhysRevC.92.054901} {\bibfield
  {journal} {\bibinfo  {journal} {Phys.\ Rev.\ C}\ }\textbf {\bibinfo {volume}
  {92}},\ \bibinfo {pages} {054901} (\bibinfo {year} {2015}{\natexlab{b}})},\
  \Eprint {http://arxiv.org/abs/1506.05763} {arXiv:1506.05763 [nucl-th]}
  \BibitemShut {NoStop}%
\bibitem [{\citenamefont {Vovchenko}\ \emph {et~al.}(2016)\citenamefont
  {Vovchenko}, \citenamefont {Poberezhnyuk}, \citenamefont {Anchishkin},\ and\
  \citenamefont {Gorenstein}}]{Vovchenko:2015uda}%
  \BibitemOpen
  \bibfield  {author} {\bibinfo {author} {\bibfnamefont {V.}~\bibnamefont
  {Vovchenko}}, \bibinfo {author} {\bibfnamefont {R.~V.}\ \bibnamefont
  {Poberezhnyuk}}, \bibinfo {author} {\bibfnamefont {D.~V.}\ \bibnamefont
  {Anchishkin}}, \ and\ \bibinfo {author} {\bibfnamefont {M.~I.}\ \bibnamefont
  {Gorenstein}},\ }\href {\doibase 10.1088/1751-8113/49/1/015003} {\bibfield
  {journal} {\bibinfo  {journal} {J. Phys. A}\ }\textbf {\bibinfo {volume}
  {49}},\ \bibinfo {pages} {015003} (\bibinfo {year} {2016})},\ \Eprint
  {http://arxiv.org/abs/1507.06537} {arXiv:1507.06537 [nucl-th]} \BibitemShut
  {NoStop}%
\bibitem [{\citenamefont {Chen}\ \emph {et~al.}(2016)\citenamefont {Chen},
  \citenamefont {Deng}, \citenamefont {Kohyama},\ and\ \citenamefont
  {Labun}}]{Chen:2015dra}%
  \BibitemOpen
  \bibfield  {author} {\bibinfo {author} {\bibfnamefont {J.-W.}\ \bibnamefont
  {Chen}}, \bibinfo {author} {\bibfnamefont {J.}~\bibnamefont {Deng}}, \bibinfo
  {author} {\bibfnamefont {H.}~\bibnamefont {Kohyama}}, \ and\ \bibinfo
  {author} {\bibfnamefont {L.}~\bibnamefont {Labun}},\ }\href {\doibase
  10.1103/PhysRevD.93.034037} {\bibfield  {journal} {\bibinfo  {journal} {Phys.
  Rev. D}\ }\textbf {\bibinfo {volume} {93}},\ \bibinfo {pages} {034037}
  (\bibinfo {year} {2016})},\ \Eprint {http://arxiv.org/abs/1509.04968}
  {arXiv:1509.04968 [hep-ph]} \BibitemShut {NoStop}%
\bibitem [{\citenamefont {Mukherjee}\ \emph {et~al.}(2017)\citenamefont
  {Mukherjee}, \citenamefont {Steinheimer},\ and\ \citenamefont
  {Schramm}}]{Mukherjee:2016nhb}%
  \BibitemOpen
  \bibfield  {author} {\bibinfo {author} {\bibfnamefont {A.}~\bibnamefont
  {Mukherjee}}, \bibinfo {author} {\bibfnamefont {J.}~\bibnamefont
  {Steinheimer}}, \ and\ \bibinfo {author} {\bibfnamefont {S.}~\bibnamefont
  {Schramm}},\ }\href {\doibase 10.1103/PhysRevC.96.025205} {\bibfield
  {journal} {\bibinfo  {journal} {Phys. Rev. C}\ }\textbf {\bibinfo {volume}
  {96}},\ \bibinfo {pages} {025205} (\bibinfo {year} {2017})},\ \Eprint
  {http://arxiv.org/abs/1611.10144} {arXiv:1611.10144 [nucl-th]} \BibitemShut
  {NoStop}%
\bibitem [{\citenamefont {Motornenko}\ \emph {et~al.}(2020)\citenamefont
  {Motornenko}, \citenamefont {Steinheimer}, \citenamefont {Vovchenko},
  \citenamefont {Schramm},\ and\ \citenamefont
  {Stoecker}}]{Motornenko:2019arp}%
  \BibitemOpen
  \bibfield  {author} {\bibinfo {author} {\bibfnamefont {A.}~\bibnamefont
  {Motornenko}}, \bibinfo {author} {\bibfnamefont {J.}~\bibnamefont
  {Steinheimer}}, \bibinfo {author} {\bibfnamefont {V.}~\bibnamefont
  {Vovchenko}}, \bibinfo {author} {\bibfnamefont {S.}~\bibnamefont {Schramm}},
  \ and\ \bibinfo {author} {\bibfnamefont {H.}~\bibnamefont {Stoecker}},\
  }\href {\doibase 10.1103/PhysRevC.101.034904} {\bibfield  {journal} {\bibinfo
   {journal} {Phys. Rev. C}\ }\textbf {\bibinfo {volume} {101}},\ \bibinfo
  {pages} {034904} (\bibinfo {year} {2020})},\ \Eprint
  {http://arxiv.org/abs/1905.00866} {arXiv:1905.00866 [hep-ph]} \BibitemShut
  {NoStop}%
\bibitem [{\citenamefont {Poberezhnyuk}\ \emph {et~al.}(2020)\citenamefont
  {Poberezhnyuk}, \citenamefont {Savchuk}, \citenamefont {Gorenstein},
  \citenamefont {Vovchenko},\ and\ \citenamefont
  {Stoecker}}]{Poberezhnyuk:2020cen}%
  \BibitemOpen
  \bibfield  {author} {\bibinfo {author} {\bibfnamefont {R.~V.}\ \bibnamefont
  {Poberezhnyuk}}, \bibinfo {author} {\bibfnamefont {O.}~\bibnamefont
  {Savchuk}}, \bibinfo {author} {\bibfnamefont {M.~I.}\ \bibnamefont
  {Gorenstein}}, \bibinfo {author} {\bibfnamefont {V.}~\bibnamefont
  {Vovchenko}}, \ and\ \bibinfo {author} {\bibfnamefont {H.}~\bibnamefont
  {Stoecker}},\ }\href@noop {} {\  (\bibinfo {year} {2020})},\ \Eprint
  {http://arxiv.org/abs/2011.06420} {arXiv:2011.06420 [hep-ph]} \BibitemShut
  {NoStop}%
\end{thebibliography}%

\end{document}